# On Huygens' Principle, Extinction Theorem, and Equivalence Principle (Metal-Material Combined System in Inhomogeneous Anisotropic Environment)

Renzun Lian

***Abstract*—In this paper, we generalize Huygens' principle (HP), extinction theorem (ET), and Franz-Harrington formulation (FHF). In our previous works, the traditional HP, ET, and FHF in homogeneous isotropic environment are generalized to inhomogeneous anisotropic lossy environment; the traditional FHF of homogeneous isotropic material system is generalized to inhomogeneous anisotropic lossy material system and then to piecewise inhomogeneous anisotropic lossy material system; the traditional HP, ET, and FHF of simply connected material system are generalized to multiply connected system and then to non-connected system; the traditional FHF of external scattering field and internal total field are generalized to internal scattering field and internal incident field. In previous work, it is proved that the generalized HP (GHP) and generalized ET (GET) are equivalent to each other; the GHP, GET, and generalized FHF (GFHF) satisfy so-called topological additivity, i.e., the GHP/GET/GFHF of whole electromagnetic (EM) system equals to the superposition of the GHP/GET/GFHF corresponding to all sub-systems.**

**In this paper, the above results obtained in our previous works, which focus on the EM system constructed by material bodies, are further generalized to the metal-material combined EM system in inhomogeneous anisotropic lossy environment, and traditional surface equivalence principle is generalized to line-surface equivalence principle.**

## I. Introduction

HUYGENS' *principle (HP)* [1], *extinction theorem (ET)* [2], and *Franz-Harrington formulation (FHF)* [3]-[5] are the important components of classical electromagnetic (EM) theory, and they have had many successful applications in EM engineering society. In paper [6], they are generalized from the following aspects:

R. Z. Lian is with the School of Electronic Science and Engineering, University of Electronic Science and Technology of China, Chengdu 611731, China. (e-mail: rzlian@vip.163.com).

■ In the aspect of EM media, the traditional HP, ET, and FHF in homogeneous isotropic environment are generalized to inhomogeneous anisotropic lossy environment; the traditional FHF of homogeneous isotropic material system is generalized to inhomogeneous anisotropic lossy material system and then to piecewise inhomogeneous anisotropic lossy material system.

■ In the aspect of topological structure of EM system, the traditional HP and ET for the case that Huygens' surface is a single closed surface is generalized to the case that "Huygens' surface" is constituted by multiple closed surfaces; the traditional FHF of a simply connected material body is generalized to a multiply connected material body and then to the EM system constructed by non-connected material bodies.

■ In the aspect of formulating fields, the traditional FHF of external scattering field and internal total field are generalized to the FHF of internal incident field and internal scattering field.

For the EM system constructed by material bodies, it is found in paper [6] that:

● The *generalized Huygens' principle (GHP)*, *generalized extinction theorem (GET)*, and *generalized Franz-Harrington formulation (GFHF)* satisfy so-called *topological additivity*, i.e., the GHP/GET/GFHF of whole EM system equals to the superposition of the GHP/GET/GFHF corresponding to all sub-systems.

● The GHP is equivalent to GET, i.e., the GHP of any field satisfies GET, and any GET corresponds to the GHP of a field.

● The GFHF of external scattering field and internal incident field is not the mathematical expression of GHP, and it is solely the summation of scattering field GHP and incident field GHP.

● The GFHF of internal total field satisfies so-called *weak extinction theorem* instead of extinction theorem. If the *piecewise Green's functions* proposed in paper [6] are utilized, the GFHF of internal total field satisfies so-called *artificial extinction theorem*, and this artificial theorem is helpful to unify the mathematical form of GFHF for various topological structures.

● The GHP is a special *surface equivalence principle (SEP)*, but SEP is not necessarily GHP. The GHP can be particularly called as *physical equivalence principle*, because it simultaneously satisfies *the concept of action at a distance*, *the law of causality*, and *the principle of superposition*. It is not necessary for SEP to simultaneously satisfy these fundamental physical requirements.

● The GFHF is not the mathematical expression of GHP and GET, and it is only the mathematical expression of SEP. The values of GFHF are mainly manifested in that various fields are uniformly expressed in terms of an identical set of equivalent surface currents, and this feature is very valuable for many engineering applications, such as solving the EM scattering and constructing the characteristic mode (CM) of material system.

In this paper, the results obtained in previous works will be further generalized to the metal-material combined system in inhomogeneous anisotropic lossy environment, and the material part of system can be any case discussed in [6]. In addition, SEP is generalized to *line-surface equivalence principle*.

This paper is organized as follows. Some necessary preparations, such as some symbols used in this paper and the topological restrictions of metallic and material parts, are provided in Sec. II. The metallic and material boundaries are decomposed in Sec. III for the preparation of decomposing various currents in Sec. IV. The currents related to metal-material combined system are decomposed in Sec. IV to reveal the dependences among them. The GHP, GET, and GFHF of metal-material combined system are provided in Sec. V, based on the Sec. IV of this paper and the results given in paper [6]. As an application, the GFHF given in Sec. V is utilized to construct the CM of metal-material combined system in Secs. VI and VII. At last, this paper is concluded in Sec. VIII.

In what follows, the $e^{j\omega t}$ convention is used throughout.

## II. PREPARATIONS

Some necessary preparations for deriving the mathematical formulation of GHP, GET, and GFHF corresponding to metal-material combined EM system are done in this section.

### A. Some symbols used in this paper

The EM system focused on by this paper is constructed by the metallic line part $L^{met}$, the metallic surface part $S^{met}$, the metallic volume part $V^{met}$, and the material body $V^{mat}$, and a typical example is shown in Fig. 1. To efficiently derive the mathematical formulation of GHP, GET, and GFHF of the structure in Fig. 1, it is necessary to employ some concepts on point set topology, such as the boundary, interior, exterior, and closure, and the rigorous mathematical definitions for them can be found in [7]. The boundaries of $L^{met}$, $S^{met}$, $V^{met}$, and $V^{mat}$ are respectively denoted as $\partial L^{met}$, $\partial S^{met}$, $\partial V^{met}$, and $\partial V^{mat}$; the interior of $V^{mat}$ is denoted as $\mathrm{int} V^{mat}$, and the exterior of $V^{mat}$ is denoted as $\mathrm{ext} V^{mat}$, i.e., $\mathrm{ext} V^{mat} \triangleq \mathbb{R}^3 \setminus \mathrm{cl} V^{mat}$, where $\mathrm{cl} V^{mat}$ represents the closure of $V^{mat}$. Obviously, both the $\mathrm{int} V^{mat}$ and $\mathrm{ext} V^{mat}$ are open sets [7].

When an external excitation $\vec{F}^{inc}$ incidents on the structure in Fig. 1, the scattering line electric current $\vec{J}^{SL}$, the scattering surface electric current $\vec{J}^{SS}_{met,surf}$, and the scattering surface electric current $\vec{J}^{SS}_{met,vol}$ will be excited on the $\partial L^{met}$, $\partial S^{met}$, and $\partial V^{met}$ respectively [8]; the scattering volume ohmic electric current $\vec{J}^{SV}_{mat,ohm}$, the scattering volume polarization electric current $\vec{J}^{SV}_{mat,pol}$, and the scattering volume magnetization magnetic current $\vec{M}^{SV}_{mat,mag}$ will be induced on the $\mathrm{int} V^{mat}$ [8], [9]. Here, the superscript "$SL$" on $\vec{J}^{SL}$ is the acronyms of term "scattering line", and the other superscripts on various currents can be similarly explained. To simplify the symbolic system of this paper, the summation of $\vec{J}^{SS}_{met,surf}$ and $\vec{J}^{SS}_{met,vol}$ is simply denoted as $\vec{J}^{SS}$, i.e., $\vec{J}^{SS} \triangleq \vec{J}^{SS}_{met,surf} + \vec{J}^{SS}_{met,vol}$; the summation of $\vec{J}^{SV}_{mat,ohm}$ and $\vec{J}^{SV}_{mat,pol}$ is simply denoted as $\vec{J}^{SV}$, i.e., $\vec{J}^{SV} \triangleq \vec{J}^{SV}_{mat,ohm} + \vec{J}^{SV}_{mat,pol}$; the $\vec{M}^{SV}_{mat,mag}$ is simply denoted as $\vec{M}^{SV}$, i.e., $\vec{M}^{SV} \triangleq \vec{M}^{SV}_{mat,mag}$.

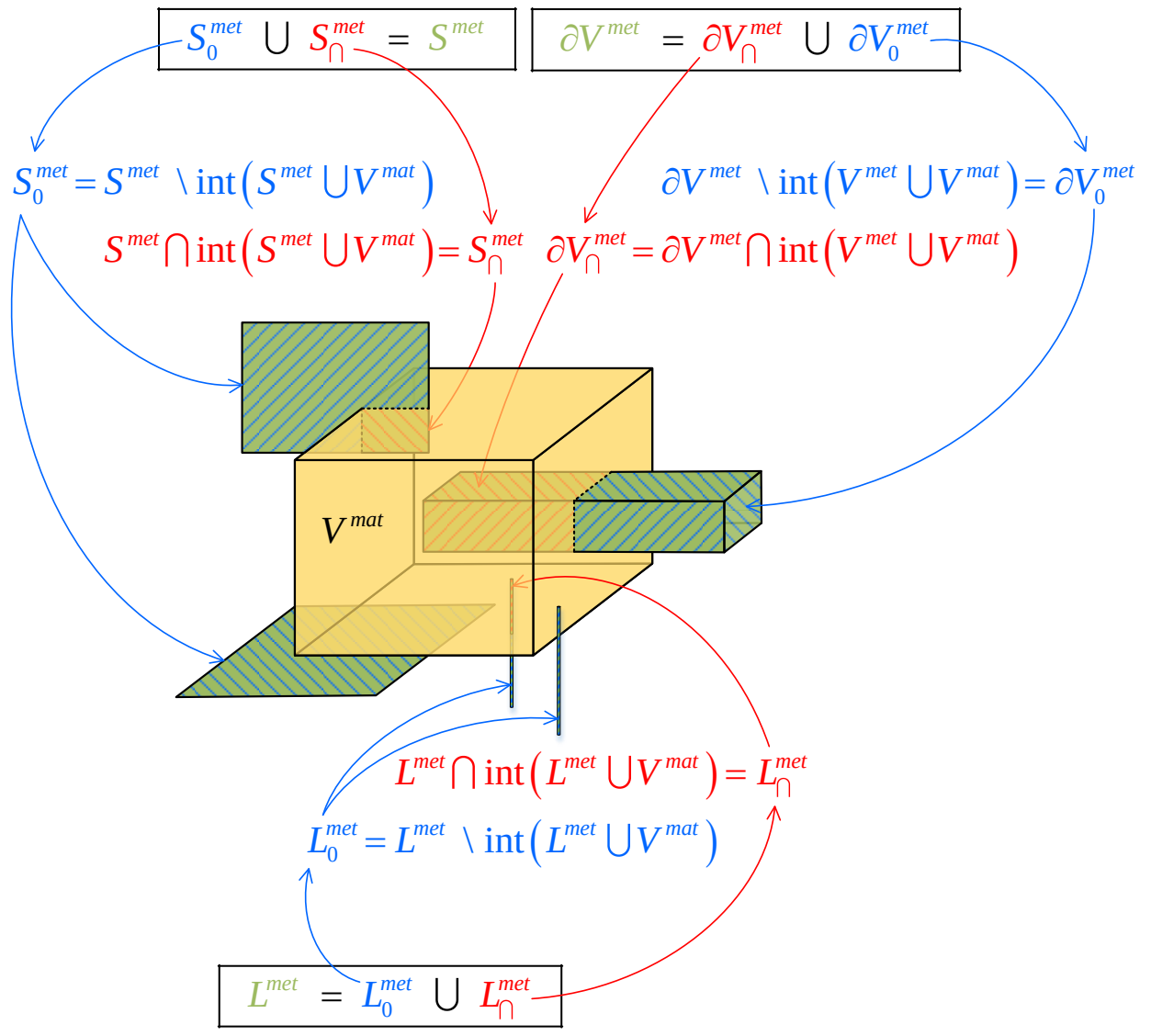

Fig. 1. The metal-material combined system considered in this paper and the decomposition for its boundary.

The scattering currents $\{\vec{J}^{SL}, \vec{J}^{SS}\}$ and $\{\vec{J}^{SV}, \vec{M}^{SV}\}$ will generate scattering field $\vec{F}^{sca}$, and the summation of $\vec{F}^{inc}$ and $\vec{F}^{sca}$ is total field $\vec{F}^{tot}$, i.e., $\vec{F}^{tot} = \vec{F}^{inc} + \vec{F}^{sca}$, where $F = E, H$. For the convenience of this paper, the $\vec{F}^{sca}$ is divided into two parts, the $\vec{F}^{sca}_{met}$ generated by metal-based scattering electric currents $\{\vec{J}^{SL}, \vec{J}^{SS}\}$ and the $\vec{F}^{sca}_{mat}$ generated by material-based scattering currents $\{\vec{J}^{SV}, \vec{M}^{SV}\}$, and $\vec{F}^{sca} = \vec{F}^{sca}_{met} + \vec{F}^{sca}_{mat}$ because of superposition principle [10].

### B. Some restrictions for the topological structure in Fig. 1, from a practical point of view

From a purely mathematical point of view, $L^{met} \subseteq \mathrm{cl} L^{met}$, and $S^{met} \subseteq \mathrm{cl} S^{met}$, and $V^{met} \subseteq \mathrm{cl} V^{met}$ [7]. However, from a practical point of view it is restricted in this paper that

$$\text{Restrction for } L^{met}: \quad L^{met} = \mathrm{cl} L^{met} \tag{1.1}$$

$$\text{Restrction for } S^{met}: \quad S^{met} = \mathrm{cl} S^{met} \tag{1.2}$$

$$\text{Restrction for } V^{met}: \quad V^{met} = \mathrm{cl} V^{met} \tag{1.3}$$

and these restrictions can be vividly understood as that there does not exist any "point-type holes" on $L^{met}$, "point-type and line-type holes" on $S^{met}$, and "point-type, line-type, and surface-type holes" on $V^{met}$. In addition, the restrictions (1.1) and (1.2) imply that $L^{met} = \partial L^{met}$ and $S^{met} = \partial S^{met}$ in $\mathbb{R}^3$ [7]. Based on the same consideration, it is also restricted that

$$\text{Restrction for } V^{mat}: \mathrm{cl} V^{mat} \setminus V^{mat} = \partial V^{mat} \cap (L^{met} \cup S^{met} \cup \partial V^{met}) \tag{1.4}$$

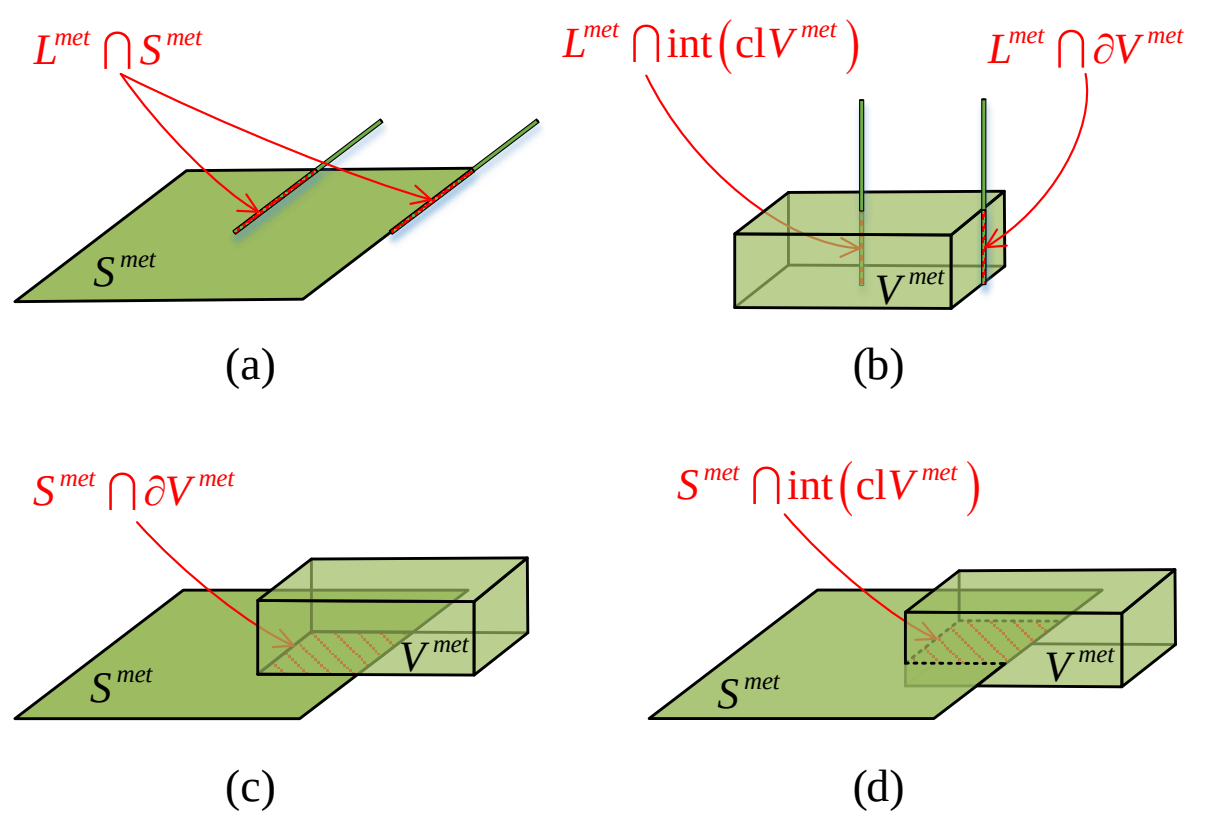

Fig. 2. (a) A part of metallic line contacts with metallic surface, and this case is not considered in this paper; (b) a part of metallic line contacts with or is submerged into metallic body, and this case is not considered in this paper; (c) a part of metallic surface contacts with metallic body, and this case is not considered in this paper; (d) a part of metallic surface is submerged into metallic body, and this case is not considered in this paper.

and this restriction can be vividly understood as that there does not exist any environment-filled "point-type, line-type, and surface-type holes" on $V^{mat}$; the "line-type holes" on $V^{mat}$ originate from the submerge of $L^{met}$ into $V^{mat}$, and the "surface-type holes" on $V^{mat}$ originate from the submergence of $S^{met}$ into $V^{mat}$. In summary, the "holes" on $V^{mat}$ are metal-filled instead of being environment-filled.

From a practical point of view, it is further restricted that

$$\text{Restrction for } L^{met}: \quad L^{met} = \mathrm{cl}\left(L^{met} \setminus \left(S^{met} \cup V^{met}\right)\right) \tag{1.1'}$$

$$\text{Restrction for } S^{met}: \quad S^{met} = \mathrm{cl}\left(S^{met} \setminus V^{met}\right). \tag{1.2'}$$

The restriction (1.1') is equivalent to saying that the intersection between $L^{met}$ and $S^{met} \cup V^{met}$ can only be some discrete points, and cannot be any lines; the restriction (1.2') is equivalent to saying that the intersection between $S^{met}$ and $V^{met}$ can only be some discrete points or lines, and cannot be any surfaces. These imply that the structures in Fig. 2 are not considered in this paper.

In addition, it is also restricted that $V^{mat}$ is a simply connected inhomogeneous anisotropic lossy material body, and that the material parameters $\vec{\vec{\sigma}}$, $\vec{\vec{\varepsilon}}$, and $\vec{\vec{\mu}}$ are two-order symmetrical tensors as explained in paper [6]. The multiply connected case, the non-connected case, and the piecewise inhomogeneous anisotropic lossy case can be similarly discussed, and corresponding mathematical formulations are formally identical to the formulations given in this paper.

## III. Boundary Decomposition

All currents appearing in the following parts of this paper distribute on the metallic and material boundaries of the structure in Fig. 1, so this section decomposes the boundaries into some sub-boundaries, to prepare for decomposing corresponding currents in the next section.

### A. *The decomposition for metallic boundary*

The $L^{met}$, $S^{met}$, and $\partial V^{met}$ can be decomposed as follows:

$$L^{met} = L_0^{met} \cup L_\cap^{met} \tag{2}$$

$$S^{met} = S_0^{met} \cup S_\cap^{met} \tag{3}$$

$$\partial V^{met} = \partial V_0^{met} \cup \partial V_\cap^{met} \tag{4}$$

where the $L_0^{met}$ and $L_\cap^{met}$ are defined as

$$L_0^{met} \triangleq L^{met} \setminus \mathrm{int}\left(L^{met} \cup V^{mat}\right) \tag{5.1}$$

$$L_\cap^{met} \triangleq L^{met} \cap \mathrm{int}\left(L^{met} \cup V^{mat}\right) \tag{5.2}$$

and the $S_0^{met}$ and $S_\cap^{met}$ are defined as

$$S_0^{met} \triangleq S^{met} \setminus \mathrm{int}\left(S^{met} \cup V^{mat}\right) \tag{6.1}$$

$$S_\cap^{met} \triangleq S^{met} \cap \mathrm{int}\left(S^{met} \cup V^{mat}\right) \tag{6.2}$$

and the $\partial V_0^{met}$ and $\partial V_\cap^{met}$ are defined as

$$\partial V_0^{met} \triangleq \partial V^{met} \setminus \mathrm{int}\left(V^{met} \cup V^{mat}\right) \tag{7.1}$$

$$\partial V_\cap^{met} \triangleq \partial V^{met} \cap \mathrm{int}\left(V^{met} \cup V^{mat}\right). \tag{7.2}$$

The $L_0^{met}$ and $L_\cap^{met}$ can be vividly understood as the part which is not submerged into $V^{mat}$ and the part which is submerged into $V^{mat}$, and the $S_0^{met}$ and $S_\cap^{met}$ can be similarly explained; the $\partial V_0^{met}$ and $\partial V_\cap^{met}$ can be vividly understood as the part which contacts with environment and the part which contacts with material body $V^{mat}$. In addition, it is obvious that

$$L_0^{met} \cap L_\cap^{met} = \varnothing \tag{8}$$

$$S_0^{met} \cap S_\cap^{met} = \varnothing \tag{9}$$

$$\partial V_0^{met} \cap \partial V_\cap^{met} = \varnothing. \tag{10}$$

### B. *The decomposition for material boundary*

As pointed out in (1), there doesn't exist any environment-filled "point-type, line-type, and surface-type holes" on $V^{mat}$, so $\partial V^{mat}$ can be decomposed into the following four parts:

$$\text{Boundary Point Part}: \partial V_{point}^{mat} = \varnothing \tag{11.1}$$

$$\text{Boundary Line Part}: \partial V_{line}^{mat} = L_\cap^{met} \tag{11.2}$$

$$\text{Boundary Open Surface Part}: \partial V_{open\ surf}^{mat} = S_\cap^{met} \tag{11.3}$$

$$\text{Boundary Closed Surface Part}: \partial V_{closed\ surf}^{mat} = \partial V^{mat} \setminus \left(L_\cap^{met} \cup S_\cap^{met}\right). \tag{11.4}$$

Obviously, the above four parts are pairwise disjoint, and

1) The boundary point part (i.e. the metal-filled "point-type holes" on $V^{mat}$) does not exist on $V^{mat}$, based on (1).

2) The boundary line part (i.e. the metal-filled "line-type holes" on $V^{mat}$) originates from the submergence of metallic lines into material body, and it is constituted by some lines only, and it does not include any surfaces and discrete points.

3) The boundary open surface part (i.e. the metal-filled "surface-type holes" on $V^{mat}$) originates from the submergence of metallic surfaces into material body, and it is constituted by some open surfaces only, and it does not include any lines, closed surfaces, and discrete points.

4) The boundary closed surface part originates from the contact between material body and environment, the contact between material body and metallic lines (here, the metallic lines are not submerged into material body), the contact between material body and metallic surfaces (here, the metallic surfaces are not submerged into material body), and the contact between material body and metallic bodies. The boundary closed surface part does not include any lines, open surfaces, and discrete points. In fact, the boundary closed surface part $\partial V_{closed\ surf}^{mat}$ can be further decomposed as follows:

$$\partial V_{closed\ surf}^{mat} = \partial V_0^{mat} \cup \partial V_{\cap}^{met} \tag{12}$$

where the $\partial V_{\cap}^{met}$ is defined as (7.2), and the $\partial V_0^{mat}$ is defined as

$$\begin{aligned} \partial V_0^{mat} &\triangleq \partial V_{closed\ surf}^{mat} \setminus \partial V_{\cap}^{met} \\ &= \left(\partial V^{mat} \setminus \left(L_{\cap}^{met} \cup S_{\cap}^{met}\right)\right) \setminus \partial V_{\cap}^{met} \\ &= \partial V^{mat} \setminus \left(L_{\cap}^{met} \cup S_{\cap}^{met} \cup \partial V_{\cap}^{met}\right) \end{aligned} \tag{13}$$

If the union of $\partial V_{open\ surf}^{mat}$ and $\partial V_{closed\ surf}^{mat}$ is denoted as $\partial V_{surf}^{mat}$ (i.e., the whole material boundary surface part is denoted as $\partial V_{surf}^{mat} \triangleq \partial V_{open\ surf}^{mat} \cup \partial V_{closed\ surf}^{mat}$), then the whole material boundary $\partial V^{mat}$ can be decomposed as follows in detail:

$$\partial V^{mat} = \overbrace{\varnothing}^{\partial V_{point}^{mat}} \cup \overbrace{L_{\cap}^{met}}^{\partial V_{line}^{mat}} \cup \overbrace{\underbrace{S_{\cap}^{met}}_{\partial V_{open\ surf}^{mat}} \cup \underbrace{\partial V_{\cap}^{met} \cup \partial V_0^{mat}}_{\partial V_{closed\ surf}^{mat}}}^{\partial V_{surf}^{mat}} . \tag{14}$$

## IV. Current Decomposition Method

Based on the boundary decomposition given in above section, the *current decomposition method* is provided in this section, and then the relationships among various sub-currents are discussed in detail for deriving GHP, GET, and GFHF in the next section.

### A. The decompositions for metal-based scattering currents

Based on (2)-(4) and (8)-(10), the scattering electric currents $\vec{J}^{SL}$ and $\vec{J}^{SS}$ can be correspondingly decomposed as follows:

$$\vec{J}^{SL}(\vec{r}) = \vec{J}_0^{SL}(\vec{r}) + \vec{J}_{\cap}^{SL}(\vec{r}) \quad , \quad \left(\vec{r} \in L^{met}\right) \tag{15}$$

$$\vec{J}^{SS}(\vec{r}) = \vec{J}_0^{SS}(\vec{r}) + \vec{J}_{\cap}^{SS}(\vec{r}) \quad , \quad \left(\vec{r} \in S^{met} \cup \partial V^{met}\right) \tag{16}$$

where the $\vec{J}_0^{SL}$ and $\vec{J}_{\cap}^{SL}$ are defined as

$$\vec{J}_0^{SL}(\vec{r}) \triangleq \begin{cases} \vec{J}^{SL}(\vec{r}) & , \quad \left(\vec{r} \in L_0^{met}\right) \\ 0 & , \quad \left(\vec{r} \in L_{\cap}^{met}\right) \end{cases} \tag{17.1}$$

$$\vec{J}_{\cap}^{SL}(\vec{r}) \triangleq \begin{cases} 0 & , \quad \left(\vec{r} \in L_0^{met}\right) \\ \vec{J}^{SL}(\vec{r}) & , \quad \left(\vec{r} \in L_{\cap}^{met}\right) \end{cases} \tag{17.2}$$

and the $\vec{J}_0^{SS}$ and $\vec{J}_{\cap}^{SS}$ are defined as

$$\vec{J}_0^{SS}(\vec{r}) \triangleq \begin{cases} \vec{J}^{SS}(\vec{r}) & , \quad \left(\vec{r} \in S_0^{met} \cup \partial V_0^{met}\right) \\ 0 & , \quad \left(\vec{r} \in S_{\cap}^{met} \cup \partial V_{\cap}^{met}\right) \end{cases} \tag{18.1}$$

$$\vec{J}_{\cap}^{SS}(\vec{r}) \triangleq \begin{cases} 0 & , \quad \left(\vec{r} \in S_0^{met} \cup \partial V_0^{met}\right) \\ \vec{J}^{SS}(\vec{r}) & , \quad \left(\vec{r} \in S_{\cap}^{met} \cup \partial V_{\cap}^{met}\right) \end{cases} . \tag{18.2}$$

### B. The decompositions for material-based equivalent currents

In this subsection, the equivalent current on whole material boundary $\partial V^{mat}$ is separately defined according to the boundary decomposition formulation (14).

**1) The equivalent surface currents on $\partial V_{closed\ surf}^{mat}$ (i.e. on $\partial V_0^{mat} \cup \partial V_{\cap}^{met}$)**

Based on papers [4]-[6], the equivalent surface currents $\{\vec{J}_{closed\ surf}^{ES}, \vec{M}_{closed\ surf}^{ES}\}$ on boundary closed surface part $\partial V_{closed\ surf}^{mat}$ are as follows:

$$\vec{J}_{closed\ surf}^{ES}(\vec{r}) = \vec{J}_0^{ES}(\vec{r}) + \vec{J}_{\partial V_{\cap}^{met}}^{ES}(\vec{r}) \quad , \quad \left(\vec{r} \in \partial V_{closed\ surf}^{mat}\right) \tag{19.1}$$

$$\vec{M}_{closed\ surf}^{ES}(\vec{r}) = \vec{M}_0^{ES}(\vec{r}) + \vec{M}_{\partial V_{\cap}^{met}}^{ES}(\vec{r}) \quad , \quad \left(\vec{r} \in \partial V_{closed\ surf}^{mat}\right) \tag{19.2}$$

in which the $\{\vec{J}_0^{ES}, \vec{M}_0^{ES}\}$ are defined as follows: [4], [5]

$$\vec{J}_0^{ES}(\vec{r}) \triangleq \hat{n}_{\rightarrow mat}(\vec{r}) \times \left[\vec{H}^{tot}(\vec{r}\,')\right]_{\vec{r}\,' \rightarrow \vec{r}} \quad , \quad \left(\vec{r} \in \partial V_0^{mat}\right) \tag{20.1}$$

$$\vec{M}_0^{ES}(\vec{r}) \triangleq \left[\vec{E}^{tot}(\vec{r}\,')\right]_{\vec{r}\,' \rightarrow \vec{r}} \times \hat{n}_{\rightarrow mat}(\vec{r}) \quad , \quad \left(\vec{r} \in \partial V_0^{mat}\right) \tag{20.2}$$

and the $\{\vec{J}_{\partial V_{\cap}^{met}}^{ES}, \vec{M}_{\partial V_{\cap}^{met}}^{ES}\}$ are defined as

$$\vec{J}_{\partial V_{\cap}^{met}}^{ES}(\vec{r}) \triangleq \hat{n}_{\rightarrow mat}(\vec{r}) \times \left[\vec{H}^{tot}(\vec{r}\,')\right]_{\vec{r}\,' \rightarrow \vec{r}} \quad , \quad \left(\vec{r} \in \partial V_{\cap}^{met}\right) \tag{21.1}$$

$$\vec{M}_{\partial V_{\cap}^{met}}^{ES}(\vec{r}) \triangleq \left[\vec{E}^{tot}(\vec{r}\,')\right]_{\vec{r}\,' \rightarrow \vec{r}} \times \hat{n}_{\rightarrow mat}(\vec{r}) \quad , \quad \left(\vec{r} \in \partial V_{\cap}^{met}\right) \tag{21.2}$$

where $\vec{r}\,' \in \text{int}V^{mat}$, and $\vec{r}\,'$ tends to $\vec{r}$ as illustrated in the subscripts in (20) and (21); $\hat{n}_{\rightarrow mat}$ is the normal vector of $\partial V_{closed\ surf}^{mat}$, and points to $\text{int}V^{mat}$. It should be emphasized that the equivalent surface currents defined in [5] equal to the $\{-\vec{J}_0^{ES}, -\vec{M}_0^{ES}\}$, because the normal vector used in [5] is $-\hat{n}_{\rightarrow mat}$ instead of $\hat{n}_{\rightarrow mat}$.

**2) The equivalent surface currents on $\partial V_{open\ surf}^{mat}$ (i.e. on $S_{\cap}^{met}$)**

To efficiently introduce the equivalent surface currents on the boundary open surface part $S_{\cap}^{met}$, we consider the example illustrated in Figs. 3 (a) and 3 (a') (i.e., a thick metallic slab $V_{slab}^{met}$ is submerged into the material body) at first, and then the $S_{\cap}^{met}$ shown in Figs. 3 (b) and 3 (b') is viewed as the limitation of $V_{slab}^{met}$ when the thickness of $V_{slab}^{met}$ tends to zero.

The plus and minus faces of $V_{slab}^{met}$ are denoted as $S_{slab;+}^{met}$ and $S_{slab;-}^{met}$ respectively, and the scattering surface electric currents on $S_{slab;+}^{met}$ and $S_{slab;-}^{met}$ are denoted as $\vec{J}_{S_{slab;+}^{met}}^{SS}$ and $\vec{J}_{S_{slab;-}^{met}}^{SS}$. Obviously, the $S_{slab;+}^{met}$ and $S_{slab;-}^{met}$ are the parts of material boundary (i.e., $S^{met}_{slab;+}, S^{met}_{slab;-} \subset \partial V^{mat}$), and the material-based equivalent surface currents on $S^{met}_{slab;+}$ and $S^{met}_{slab;-}$ are denoted as $\{\vec{J}^{ES}_{S^{met}_{slab;+}}, \vec{M}^{ES}_{S^{met}_{slab;+}}\}$ and $\{\vec{J}^{ES}_{S^{met}_{slab;-}}, \vec{M}^{ES}_{S^{met}_{slab;-}}\}$ respectively. If the thickness of metallic slab is denoted as $D^{met}_{slab}$, the following limitations exist:

$$\lim_{D^{met}_{slab} \to 0} S^{met}_{slab;\pm} = S^{met}_{\cap} \tag{22}$$

$$\lim_{D^{met}_{slab} \to 0} \left( \vec{J}^{SS}_{S^{met}_{slab;+}} + \vec{J}^{SS}_{S^{met}_{slab;-}} \right) = \vec{J}^{SS}_{\cap} \tag{23}$$

$$\lim_{D^{met}_{slab} \to 0} \vec{J}^{ES}_{S^{met}_{slab;\pm}} = \vec{J}^{ES}_{\pm} \tag{24.1}$$

$$\lim_{D^{met}_{slab} \to 0} \vec{M}^{ES}_{S^{met}_{slab;\pm}} = \vec{M}^{ES}_{\pm} \tag{24.2}$$

where the $\{\vec{J}^{ES}_{\pm}, \vec{M}^{ES}_{\pm}\}$ on $S^{met}_{\cap}$ are defined as follows:

$$\vec{J}^{ES}_{\pm}(\vec{r}) \triangleq \hat{n}_{\pm}(\vec{r}) \times \left[ \vec{H}^{tot}(\vec{r}_{\pm}) \right]_{\vec{r}_{\pm} \to \vec{r}} \quad , \quad (\vec{r} \in S^{met}_{\cap}) \tag{25.1}$$

$$\vec{M}^{ES}_{\pm}(\vec{r}) \triangleq \left[ \vec{E}^{tot}(\vec{r}_{\pm}) \right]_{\vec{r}_{\pm} \to \vec{r}} \times \hat{n}_{\pm}(\vec{r}) \quad , \quad (\vec{r} \in S^{met}_{\cap}) . \tag{25.2}$$

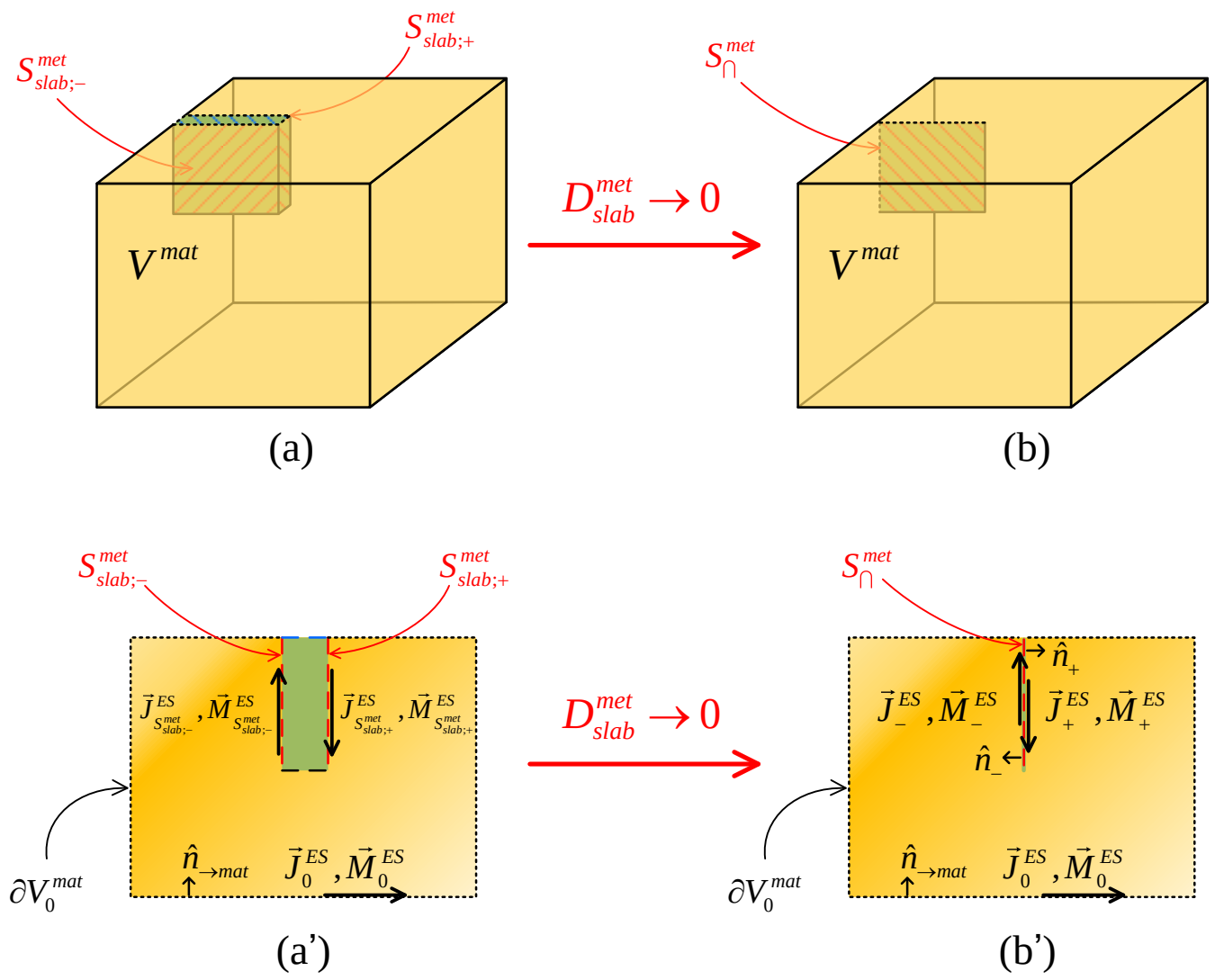

Fig. 3. (a) A thick metallic slab is submerged into material body; (b) a metallic surface is submerged into material body; (a') the sectional view of Fig. 3 (a), and the equivalent surface currents $\{\vec{J}^{ES}_{S^{met}_{slab;+}}, \vec{M}^{ES}_{S^{met}_{slab;+}}\}$ and $\{\vec{J}^{ES}_{S^{met}_{slab;-}}, \vec{M}^{ES}_{S^{met}_{slab;-}}\}$; (b') the sectional view of Fig. 3 (b), and the equivalent surface currents $\{\vec{J}^{ES}_+, \vec{M}^{ES}_+\}$ and $\{\vec{J}^{ES}_-, \vec{M}^{ES}_-\}$.

In (25), $\vec{r}_+, \vec{r}_- \in \text{int} V^{mat}$; $\vec{r}_+$ and $\vec{r}_-$ tend to $\vec{r}$ from the plus and minus sides of $S^{met}_{\cap}$ respectively; $\hat{n}_+$ and $\hat{n}_-$ are the normal vectors of $S^{met}_{\cap}$, and they point to the plus and minus sides of $S^{met}_{\cap}$ respectively.

Because of superposition principle [10], the summation of the fields respectively generated by $\{\vec{J}^{ES}_+, \vec{M}^{ES}_+\}$ and $\{\vec{J}^{ES}_-, \vec{M}^{ES}_-\}$ are identical to the field generated by $\{\vec{J}^{ES}_+ + \vec{J}^{ES}_-, \vec{M}^{ES}_+ + \vec{M}^{ES}_-\}$, and then the $\{\vec{J}^{ES}_+ + \vec{J}^{ES}_-, \vec{M}^{ES}_+ + \vec{M}^{ES}_-\}$ is treated as a whole in this paper. In addition, considering of that both the domain of $\{\vec{J}^{ES}_+, \vec{M}^{ES}_+\}$ and the domain of $\{\vec{J}^{ES}_-, \vec{M}^{ES}_-\}$ are $S^{met}_{\cap}$ and that $\hat{n}_-(\vec{r}) = -\hat{n}_+(\vec{r})$ for any $\vec{r} \in S^{met}_{\cap}$, the equivalent surface currents on the boundary open surface part $S^{met}_{\cap}$ can be defined as

$$\begin{aligned} \vec{J}^{ES}_{open\ surf}(\vec{r}) &\triangleq \vec{J}^{ES}_+(\vec{r}) + \vec{J}^{ES}_-(\vec{r}) \\ &= \hat{n}_+(\vec{r}) \times \left[ \vec{H}^{tot}(\vec{r}_+) - \vec{H}^{tot}(\vec{r}_-) \right]_{\vec{r}_+, \vec{r}_- \to \vec{r}} \quad , \quad (\vec{r} \in S^{met}_{\cap}) \end{aligned} \tag{26.1}$$

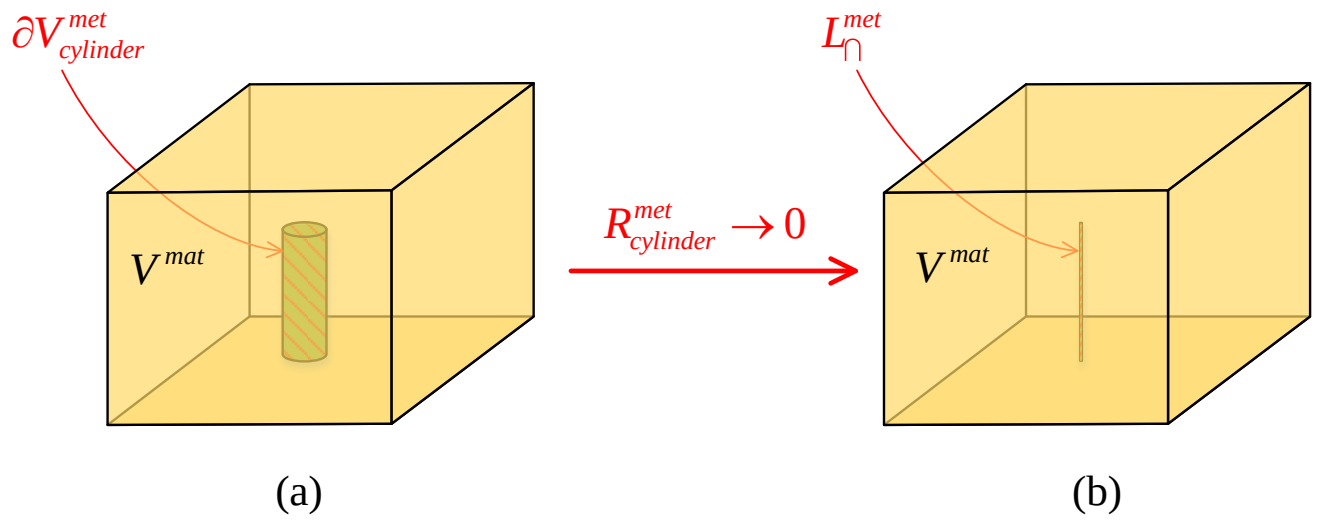

Fig. 4. (a) A metallic cylinder is submerged into material body; (b) a metallic line is submerged into material body.

$$\begin{aligned} \vec{M}^{ES}_{open\ surf}(\vec{r}) &\triangleq \vec{M}^{ES}_+(\vec{r}) + \vec{M}^{ES}_-(\vec{r}) \\ &= \left[ \vec{E}^{tot}(\vec{r}_+) - \vec{E}^{tot}(\vec{r}_-) \right]_{\vec{r}_+, \vec{r}_- \to \vec{r}} \times \hat{n}_+(\vec{r}) \quad , \quad (\vec{r} \in S^{met}_{\cap}) . \end{aligned} \tag{26.2}$$

**3) The equivalent line currents on $\partial V^{mat}_{line}$ (i.e. on $L^{met}_{\cap}$)**

To efficiently introduce the equivalent line currents $\{\vec{J}^{EL}, \vec{M}^{EL}\}$ on the boundary line part $L^{met}_{\cap}$, we consider the example illustrated in Fig. 4 (a) (i.e., a metallic cylinder $V^{met}_{cylinder}$ is submerged into the material body) at first, and then the $L^{met}_{\cap}$ shown in Fig. 4 (b) is viewed as the limitation of $V^{met}_{cylinder}$ when the radius of $V^{met}_{cylinder}$ tends to zero.

The boundary of $V^{met}_{cylinder}$ is denoted as $\partial V^{met}_{cylinder}$, and the scattering surface electric current on $\partial V^{met}_{cylinder}$ is denoted as $\vec{J}^{SS}_{\partial V^{met}_{cylinder}}$. Obviously, the $\partial V^{met}_{cylinder}$ is a part of material boundary (i.e., $\partial V^{met}_{cylinder} \subset \partial V^{mat}$), and the material-based equivalent surface currents on $\partial V^{met}_{cylinder}$ are denoted as $\{\vec{J}^{ES}_{\partial V^{met}_{cylinder}}, \vec{M}^{ES}_{\partial V^{met}_{cylinder}}\}$. If the radius of $V^{met}_{cylinder}$ is denoted as $R^{met}_{cylinder}$, the following limitations exist:

$$\partial V^{met}_{cylinder} \xrightarrow{R^{met}_{cylinder} \to 0} L^{met}_{\cap} \tag{27}$$

$$\vec{J}^{SS}_{\partial V^{met}_{cylinder}} \xrightarrow{R^{met}_{cylinder} \to 0} \vec{J}^{SL}_{\cap} \tag{28}$$

$$\vec{J}^{ES}_{\partial V^{met}_{cylinder}} \xrightarrow{R^{met}_{cylinder} \to 0} \vec{J}^{EL} \tag{29.1}$$

$$\vec{M}^{ES}_{\partial V^{met}_{cylinder}} \xrightarrow{R^{met}_{cylinder} \to 0} \vec{M}^{EL} \tag{29.2}$$

and then the equivalent line currents $\{\vec{J}^{EL}, \vec{M}^{EL}\}$ on the boundary line part $L^{met}_{\cap}$ can be defined as follows:

$$\vec{J}^{EL}(\vec{r}) \triangleq \hat{e}_l \lim_{\vec{r}' \to \vec{r}} \oint_{C(\vec{r}')} \vec{H}^{tot}(\vec{r}') \cdot d\vec{l}' \quad , \quad (\vec{r} \in L^{met}_{\cap}) \tag{30.1}$$

$$\vec{M}^{EL}(\vec{r}) \triangleq -\hat{e}_l \lim_{\vec{r}' \to \vec{r}} \oint_{C(\vec{r}')} \vec{E}^{tot}(\vec{r}') \cdot d\vec{l}' \quad , \quad (\vec{r} \in L^{met}_{\cap}) . \tag{30.2}$$

In (30), the integral path $C(\vec{r}')$ is a circle constructed by the points $\vec{r}'$ which are in the set $\text{int} V^{mat}$ and tend to point $\vec{r}$; $\hat{e}_l$ is the reference direction of $\vec{J}^{EL}$ and $\vec{M}^{EL}$; the $\hat{e}_l$ and the reference direction of $C(\vec{r}')$ satisfy right-hand rule.

**4) Summary**

In summary, the whole material boundary $\partial V^{mat}$ can be decomposed into four parts as (11) or more elaborately decomposed into five parts as (14), and then the equivalent currents on $\partial V^{mat}$ can be correspondingly defined as (20), (21), (26), and (30). To simplify the symbolic system of the following parts of this paper, the summation of $\vec{C}^{ES}_{\partial V^{met}_{\cap}}$ and $\vec{C}^{ES}_{open\ surf}$ is denoted as $\vec{C}^{ES}_{\cap}$ (because $\vec{C}^{ES}_{\partial V^{met}_{\cap}}$ and $\vec{C}^{ES}_{open\ surf}$ exist on the intersection be-tween $\partial V^{mat}$ and $\partial V_{\cap}^{met} \cup S_{\cap}^{met}$), and the summation of $\vec{C}_{closed\ surf}^{ES}$ and $\vec{C}_{open\ surf}^{ES}$ is denoted as $\vec{C}^{ES}$ (because $\vec{C}_{closed\ surf}^{ES}$ and $\vec{C}_{open\ surf}^{ES}$ constitute the whole of equivalent surface currents), i.e.,

$$\text{Equivalent Currents on } \partial V^{mat}: \left\{ \vec{C}^{EL}, \overbrace{\overbrace{\vec{C}_0^{ES}, \vec{C}_{\partial V_{\cap}^{met}}^{ES}}^{\vec{C}_{closed\ surf}^{ES}}, \overbrace{\vec{C}_{+}^{ES}, \vec{C}_{-}^{ES}}^{\vec{C}_{open\ surf}^{ES}}}^{\vec{C}^{ES}} \right\} \quad (31)$$

(with $\vec{C}_{\partial V_{\cap}^{met}}^{ES}, \vec{C}_{+}^{ES}, \vec{C}_{-}^{ES}$ grouped under $\vec{C}_{\cap}^{ES}$)

where $C = J, M$.

*C. The relationships among various sub-currents*

Based on the above discussions, all the sub-currents on the boundary of a metal-material combined system are as follows:

$$\text{Electric Currents}: \left\{ \overbrace{\overbrace{\vec{J}_0^{SL}, \vec{J}_{\cap}^{SL}}^{\vec{J}^{SL}}, \overbrace{\vec{J}_0^{SS}, \vec{J}_{\cap}^{SS}}^{\vec{J}^{SS}}}^{\text{scattering } \vec{J} \text{ on metal boundary}}, \overbrace{\vec{J}^{EL}, \underbrace{\overbrace{\vec{J}_0^{ES}, \vec{J}_{\partial V_{\cap}^{met}}^{ES}}^{\vec{J}_{closed\ surf}^{ES}}, \overbrace{\vec{J}_{+}^{ES}, \vec{J}_{-}^{ES}}^{\vec{J}_{open\ surf}^{ES}}}_{\vec{J}^{ES}}}^{\text{equivalent } \vec{J} \text{ on material boundary}} \right\} \quad (32.1)$$

(with $\vec{J}_{\partial V_{\cap}^{met}}^{ES}, \vec{J}_{+}^{ES}, \vec{J}_{-}^{ES}$ grouped under $\vec{J}_{\cap}^{ES}$)

$$\text{Magnetic Currents}: \left\{ \overbrace{\vec{M}^{EL}, \underbrace{\overbrace{\vec{M}_0^{ES}, \vec{M}_{\partial V_{\cap}^{met}}^{ES}}^{\vec{M}_{closed\ surf}^{ES}}, \overbrace{\vec{M}_{+}^{ES}, \vec{M}_{-}^{ES}}^{\vec{M}_{open\ surf}^{ES}}}_{\vec{M}^{ES}}}^{\text{equivalent } \vec{M} \text{ on material boundary}} \right\}. \quad (32.2)$$

(with $\vec{M}_{\partial V_{\cap}^{met}}^{ES}, \vec{M}_{+}^{ES}, \vec{M}_{-}^{ES}$ grouped under $\vec{M}_{\cap}^{ES}$)

Due to the tangential boundary conditions of $\vec{H}^{tot}$ and $\vec{E}^{tot}$ on $\partial V_{\cap}^{met}$, it is easy to derive that

$$\vec{J}_{\partial V_{\cap}^{met}}^{ES}(\vec{r}) = \vec{J}_{\partial V_{\cap}^{met}}^{SS}(\vec{r}), \quad (\vec{r} \in \partial V_{\cap}^{met}) \quad (33.1)$$

$$\vec{M}_{\partial V_{\cap}^{met}}^{ES}(\vec{r}) = 0, \quad (\vec{r} \in \partial V_{\cap}^{met}). \quad (33.2)$$

Due to the same reasons as deriving (33), the following relations for the currents defined in Sec. IV-B 2) can be derived:

$$\vec{J}_{S_{slab;\pm}^{met}}^{ES}(\vec{r}) = \vec{J}_{S_{slab;\pm}^{met}}^{SS}(\vec{r}), \quad (\vec{r} \in S_{slab;\pm}^{met}) \quad (34.1)$$

$$\vec{M}_{S_{slab;\pm}^{met}}^{ES}(\vec{r}) = 0, \quad (\vec{r} \in S_{slab;\pm}^{met}) \quad (34.2)$$

and then

$$\vec{J}_{open\ surf}^{ES}(\vec{r}) = \vec{J}_{\cap}^{SS}(\vec{r}), \quad (\vec{r} \in S_{\cap}^{met}) \quad (35.1)$$

$$\vec{M}_{open\ surf}^{ES}(\vec{r}) = 0, \quad (\vec{r} \in S_{\cap}^{met}) \quad (35.2)$$

because of (23), (24), and (26).

In fact, the above (33) and (35) can be uniformly written as follows:

$$\vec{J}_{\cap}^{SS}(\vec{r}) = \vec{J}_{\cap}^{ES}(\vec{r}) = \begin{cases} \vec{J}_{open\ surf}^{ES}(\vec{r}), & (\vec{r} \in S_{\cap}^{met}) \\ \vec{J}_{\partial V_{\cap}^{met}}^{ES}(\vec{r}), & (\vec{r} \in \partial V_{\cap}^{met}) \end{cases} \quad (36.1)$$

$$0 = \vec{M}_{\cap}^{ES}(\vec{r}) = \begin{cases} \vec{M}_{open\ surf}^{ES}(\vec{r}), & (\vec{r} \in S_{\cap}^{met}) \\ \vec{M}_{\partial V_{\cap}^{met}}^{ES}(\vec{r}), & (\vec{r} \in \partial V_{\cap}^{met}) \end{cases}. \quad (36.2)$$

In addition, it must be EMPHASIZED that: $\vec{M}_0^{ES}(\vec{r}) = 0$ for any $\vec{r} \in \partial V_0^{mat} \cap S^{met}$, because of the tangential boundary condition of the total electric field on $S^{met}$; $[\hat{n}_{\rightarrow mat}(\vec{r}) \times \hat{e}_l(\vec{r})] \cdot \vec{M}_0^{ES}(\vec{r}) = 0$ for any $\vec{r} \in \partial V_0^{mat} \cap L^{met}$, because of the tangential boundary condition of the total electric field on $L^{met}$.

Similarly, the following relationships among the currents defined in Sec. IV-B 3) can be derived:

$$\vec{J}_{\partial V_{cylinder}^{met}}^{ES}(\vec{r}) = \vec{J}_{\partial V_{cylinder}^{met}}^{SS}(\vec{r}), \quad (\vec{r} \in \partial V_{cylinder}^{met}) \quad (37.1)$$

$$\vec{M}_{\partial V_{cylinder}^{met}}^{ES}(\vec{r}) = 0, \quad (\vec{r} \in \partial V_{cylinder}^{met}) \quad (37.2)$$

and then

$$\vec{J}^{EL}(\vec{r}) = \vec{J}_{\cap}^{SL}(\vec{r}), \quad (\vec{r} \in L_{\cap}^{met}) \quad (38.1)$$

$$\vec{M}^{EL}(\vec{r}) = 0, \quad (\vec{r} \in L_{\cap}^{met}) \quad (38.2)$$

because of (28) and (29).

The equivalent currents appeared in the material boundary of metal-material combined system include equivalent line electric current besides traditional equivalent surface electric and magnetic currents, so the corresponding equivalence principle is particularly called as *line-surface equivalence principle (LSEP)* to be distinguished from traditional SEP. In paper [6], it is found that the GFHF of material system is the mathematical expression of SEP, and the mathematical expression of the LSEP of metal-material combined system will be explicitly provided in the following Sec. V.

## V. Generalized Huygens' Principle, Extinction Theorem, and Franz-Harrington Formulation

In this section, the formulations and conclusions obtained in paper [6] for material system will be further generalized to the metal-material combined system.

The domain occupied by whole metal-material combined system is denoted as $D^{sys}$, i.e.,

$$D^{sys} \triangleq L^{met} \cup S^{met} \cup V^{met} \cup V^{mat} \quad (39)$$

and then [7]

$$\partial D^{sys} = L_0^{met} \cup S_0^{met} \cup \partial V_0^{met} \cup \partial V_0^{mat} \quad (40.1)$$

$$\text{int}\, D^{sys} = L_{\cap}^{met} \cup S_{\cap}^{met} \cup \partial V_{\cap}^{met} \cup \text{int}\, V^{met} \cup \text{int}\, V^{mat} \quad (40.2)$$

$$\begin{aligned} \text{ext}\, D^{sys} &= \mathbb{R}^3 \setminus \text{cl}\, D \\ &= \mathbb{R}^3 \setminus D \\ &= \mathbb{R}^3 \setminus \left( L^{met} \cup S^{met} \cup V^{met} \cup V^{mat} \right) \end{aligned} . \quad (40.3)$$

The second and third equalities in (40.3) are based on (1) and (39).

*A. Generalized Huygens' principle and extinction theorem*

For material system, the Huygens' surface supporting *Huygens' secondary source* is not unique, and the boundary of real source is a natural and the smallest one [6], so the Huygens' surfaces used in [6] are selected as material boundaries. However, the source boundaries of metal-material combined system are not restricted to surfaces as illustrated in Sec. III, so they are particularly called as *"Huygens' boundaries"* to be distinguished from traditional Huygens' surfaces.

**1) The GHP and GET corresponding to the "Huygens' boundary" which is selected as material boundary**

If the "Huygens' boundary" is selected as whole material boundary $\partial V^{mat}$, the incident field GHP can be mathematically written as (41.1) based on the conclusion given in paper [6].

$$\left.\begin{array}{lcc} \text{ext}\,D^{sys} & : & 0 \\ \text{int}\,V^{mat} & : & \vec{F}^{inc} \\ \text{int}\,V^{met} & : & 0 \end{array}\right\} \begin{aligned} = & \Big[\overset{\leftrightarrow}{G}{}^{JF}_{env}(\vec{r},\vec{r}\,')*\hat{e}_l \lim_{\vec{r}\,''\to\vec{r}\,'}\oint_{C(\vec{r}\,'')}\vec{H}^{inc}(\vec{r}\,'')\cdot d\vec{l}\,''\Big]_{L^{met}_{\cap}} \\ & -\Big[\overset{\leftrightarrow}{G}{}^{MF}_{env}(\vec{r},\vec{r}\,')*\hat{e}_l \lim_{\vec{r}\,''\to\vec{r}\,'}\oint_{C(\vec{r}\,'')}\vec{E}^{inc}(\vec{r}\,'')\cdot d\vec{l}\,''\Big]_{L^{met}_{\cap}} \\ & +\Big[\overset{\leftrightarrow}{G}{}^{JF}_{env}*\big(\hat{n}_{S^{met}_{\cap+}}\times\vec{H}^{inc}\big)+\overset{\leftrightarrow}{G}{}^{MF}_{env}*\big(\vec{E}^{inc}\times\hat{n}_{S^{met}_{\cap+}}\big)\Big]_{S^{met}_{\cap+}} \\ & +\Big[\overset{\leftrightarrow}{G}{}^{JF}_{env}*\big(\hat{n}_{S^{met}_{\cap-}}\times\vec{H}^{inc}\big)+\overset{\leftrightarrow}{G}{}^{MF}_{env}*\big(\vec{E}^{inc}\times\hat{n}_{S^{met}_{\cap-}}\big)\Big]_{S^{met}_{\cap-}} \\ & +\Big[\overset{\leftrightarrow}{G}{}^{JF}_{env}*\big(\hat{n}_{V^{mat}_{-}}\times\vec{H}^{inc}\big)+\overset{\leftrightarrow}{G}{}^{MF}_{env}*\big(\vec{E}^{inc}\times\hat{n}_{V^{mat}_{-}}\big)\Big]_{\partial V^{met}_{\cap+}} \\ & +\Big[\overset{\leftrightarrow}{G}{}^{JF}_{env}*\big(\hat{n}_{V^{mat}_{-}}\times\vec{H}^{inc}\big)+\overset{\leftrightarrow}{G}{}^{MF}_{env}*\big(\vec{E}^{inc}\times\hat{n}_{V^{mat}_{-}}\big)\Big]_{\partial V^{mat}_{0-}} \\ = & \Big[\overset{\leftrightarrow}{G}{}^{JF}_{env}*\big(\hat{n}_{V^{mat}_{-}}\times\vec{H}^{inc}\big)+\overset{\leftrightarrow}{G}{}^{MF}_{env}*\big(\vec{E}^{inc}\times\hat{n}_{V^{mat}_{-}}\big)\Big]_{\partial V^{met}_{\cap}} \\ & +\Big[\overset{\leftrightarrow}{G}{}^{JF}_{env}*\big(\hat{n}_{V^{mat}_{-}}\times\vec{H}^{inc}\big)+\overset{\leftrightarrow}{G}{}^{MF}_{env}*\big(\vec{E}^{inc}\times\hat{n}_{V^{mat}_{-}}\big)\Big]_{\partial V^{mat}_{0}} \end{aligned} \tag{41.1}$$

where the second equality is because of that the incident sources don't distribute on $L^{met}_{\cap}$ and $S^{met}_{\cap}$; the mathematical formulation of the GHP corresponding to material scattering field is as follows:

$$\left.\begin{array}{lcc} \text{ext}\,D^{sys} & : & \vec{F}^{sca}_{mat} \\ \text{int}\,V^{mat} & : & 0 \\ \text{int}\,V^{met} & : & \vec{F}^{sca}_{mat} \end{array}\right\} \begin{aligned} = & -\Big[\overset{\leftrightarrow}{G}{}^{JF}_{env}(\vec{r},\vec{r}\,')*\hat{e}_l \lim_{\vec{r}\,''\to\vec{r}\,'}\oint_{C(\vec{r}\,'')}\vec{H}^{sca}_{mat}(\vec{r}\,'')\cdot d\vec{l}\,''\Big]_{L^{met}_{\cap}} \\ & +\Big[\overset{\leftrightarrow}{G}{}^{MF}_{env}(\vec{r},\vec{r}\,')*\hat{e}_l \lim_{\vec{r}\,''\to\vec{r}\,'}\oint_{C(\vec{r}\,'')}\vec{E}^{sca}_{mat}(\vec{r}\,'')\cdot d\vec{l}\,''\Big]_{L^{met}_{\cap}} \\ & +\Big[\overset{\leftrightarrow}{G}{}^{JF}_{env}*\big(\hat{n}_{S^{met}_{\cap-}}\times\vec{H}^{sca}_{mat}\big)+\overset{\leftrightarrow}{G}{}^{MF}_{env}*\big(\vec{E}^{sca}_{mat}\times\hat{n}_{S^{met}_{\cap-}}\big)\Big]_{S^{met}_{\cap+}} \\ & +\Big[\overset{\leftrightarrow}{G}{}^{JF}_{env}*\big(\hat{n}_{S^{met}_{\cap+}}\times\vec{H}^{sca}_{mat}\big)+\overset{\leftrightarrow}{G}{}^{MF}_{env}*\big(\vec{E}^{sca}_{mat}\times\hat{n}_{S^{met}_{\cap+}}\big)\Big]_{S^{met}_{\cap-}} \\ & +\Big[\overset{\leftrightarrow}{G}{}^{JF}_{env}*\big(\hat{n}_{V^{mat}_{+}}\times\vec{H}^{sca}_{mat}\big)+\overset{\leftrightarrow}{G}{}^{MF}_{env}*\big(\vec{E}^{sca}_{mat}\times\hat{n}_{V^{mat}_{+}}\big)\Big]_{\partial V^{met}_{\cap-}} \\ & +\Big[\overset{\leftrightarrow}{G}{}^{JF}_{env}*\big(\hat{n}_{V^{mat}_{+}}\times\vec{H}^{sca}_{mat}\big)+\overset{\leftrightarrow}{G}{}^{MF}_{env}*\big(\vec{E}^{sca}_{mat}\times\hat{n}_{V^{mat}_{+}}\big)\Big]_{\partial V^{mat}_{0+}} \\ = & \Big[\overset{\leftrightarrow}{G}{}^{JF}_{env}*\big(\hat{n}_{V^{mat}_{+}}\times\vec{H}^{sca}_{mat}\big)+\overset{\leftrightarrow}{G}{}^{MF}_{env}*\big(\vec{E}^{sca}_{mat}\times\hat{n}_{V^{mat}_{+}}\big)\Big]_{\partial V^{met}_{\cap}} \\ & +\Big[\overset{\leftrightarrow}{G}{}^{JF}_{env}*\big(\hat{n}_{V^{mat}_{+}}\times\vec{H}^{sca}_{mat}\big)+\overset{\leftrightarrow}{G}{}^{MF}_{env}*\big(\vec{E}^{sca}_{mat}\times\hat{n}_{V^{mat}_{+}}\big)\Big]_{\partial V^{mat}_{0}} \end{aligned} \tag{42.1}$$

where the second equality is because of that there doesn't exist material-based scattering current distributing on $L^{met}_{\cap}$, $S^{met}_{\cap}$, and $\partial V^{mat}$, and this conclusion can be strictly proven by employing the method given in paper [9]; the mathematical formulation of the GHP corresponding to the metallic scattering field is as follows:

$$\left.\begin{array}{lcc} \text{ext}\,D^{sys} & : & 0 \\ \text{int}\,V^{mat} & : & \vec{F}^{sca}_{met} \\ \text{int}\,V^{met} & : & 0 \end{array}\right\} \begin{aligned} = & \Big[\overset{\leftrightarrow}{G}{}^{JF}_{env}(\vec{r},\vec{r}\,')*\hat{e}_l \lim_{\vec{r}\,''\to\vec{r}\,'}\oint_{C(\vec{r}\,'')}\vec{H}^{sca}_{met}(\vec{r}\,'')\cdot d\vec{l}\,''\Big]_{L^{met}_{\cap}} \\ & -\Big[\overset{\leftrightarrow}{G}{}^{MF}_{env}(\vec{r},\vec{r}\,')*\hat{e}_l \lim_{\vec{r}\,''\to\vec{r}\,'}\oint_{C(\vec{r}\,'')}\vec{E}^{sca}_{met}(\vec{r}\,'')\cdot d\vec{l}\,''\Big]_{L^{met}_{\cap}} \\ & +\Big[\overset{\leftrightarrow}{G}{}^{JF}_{env}*\big(\hat{n}_{S^{met}_{\cap+}}\times\vec{H}^{sca}_{met}\big)+\overset{\leftrightarrow}{G}{}^{MF}_{env}*\big(\vec{E}^{sca}_{met}\times\hat{n}_{S^{met}_{\cap+}}\big)\Big]_{S^{met}_{\cap+}} \\ & +\Big[\overset{\leftrightarrow}{G}{}^{JF}_{env}*\big(\hat{n}_{S^{met}_{\cap-}}\times\vec{H}^{sca}_{met}\big)+\overset{\leftrightarrow}{G}{}^{MF}_{env}*\big(\vec{E}^{sca}_{met}\times\hat{n}_{S^{met}_{\cap-}}\big)\Big]_{S^{met}_{\cap-}} \\ & +\Big[\overset{\leftrightarrow}{G}{}^{JF}_{env}*\big(\hat{n}_{V^{mat}_{-}}\times\vec{H}^{sca}_{met}\big)+\overset{\leftrightarrow}{G}{}^{MF}_{env}*\big(\vec{E}^{sca}_{met}\times\hat{n}_{V^{mat}_{-}}\big)\Big]_{\partial V^{met}_{\cap+}} \\ & +\Big[\overset{\leftrightarrow}{G}{}^{JF}_{env}*\big(\hat{n}_{V^{mat}_{-}}\times\vec{H}^{sca}_{met}\big)+\overset{\leftrightarrow}{G}{}^{MF}_{env}*\big(\vec{E}^{sca}_{met}\times\hat{n}_{V^{mat}_{-}}\big)\Big]_{\partial V^{mat}_{0-}} \\ = & \Big[\overset{\leftrightarrow}{G}{}^{JF}_{env}*\vec{J}^{SL}_{\cap}\Big]_{L^{met}_{\cap}}+\Big[\overset{\leftrightarrow}{G}{}^{JF}_{env}*\vec{J}^{SS}_{\cap}\Big]_{S^{met}_{\cap}} \\ & +\Big[\overset{\leftrightarrow}{G}{}^{JF}_{env}*\big(\hat{n}_{V^{mat}_{-}}\times\vec{H}^{sca}_{met}\big)+\overset{\leftrightarrow}{G}{}^{MF}_{env}*\big(\vec{E}^{sca}_{met}\times\hat{n}_{V^{mat}_{-}}\big)\Big]_{\partial V^{met}_{\cap+}} \\ & +\Big[\overset{\leftrightarrow}{G}{}^{JF}_{env}*\big(\hat{n}_{V^{mat}_{-}}\times\vec{H}^{sca}_{met}\big)+\overset{\leftrightarrow}{G}{}^{MF}_{env}*\big(\vec{E}^{sca}_{met}\times\hat{n}_{V^{mat}_{-}}\big)\Big]_{\partial V^{mat}_{0-}} \end{aligned} \tag{42.2}$$

where the second equality is because of that there doesn't exist any metal-based scattering magnetic currents on the $L^{met}_{\cap}$ and $S^{met}_{\cap}$.

In (41.1), the integral domains $S^{met}_{\cap+}$ and $S^{met}_{\cap-}$ respectively represent the plus and minus sides of $S^{met}_{\cap}$; the $\hat{n}_{S^{met}_{\cap+}}$ and $\hat{n}_{S^{met}_{\cap-}}$ are the normal vectors of $S^{met}_{\cap}$, and respectively point to the plus and minus sides of $S^{met}_{\cap}$; the integral domains $\partial V^{met}_{\cap+}$ and $\partial V^{mat}_{0-}$ respectively represent the metallic outer boundary corresponding to $\partial V^{met}_{\cap}$ and the material inner boundary corresponding to $\partial V^{mat}_{0}$; the $\hat{n}_{V^{mat}_{-}}$ is the normal vector of $\partial V^{mat}$, and points to the interior of $V^{mat}$; the various Green's functions are the environment Green's functions used in paper [6]. The other symbols in (41.1), (42.1), and (42.2) can be similarly explained.

**2) The GHP and GET corresponding to the "Huygens' boundary" which is selected as metallic boundary**

If the "Huygens' boundary" is selected as whole metallic boundary $L^{met}\bigcup S^{met}\bigcup\partial V^{met}$, the incident field GHP can be mathematically written as the following (41.2) based on the conclusion given in paper [6]:

$$\left.\begin{array}{lcc} \text{ext}\,D^{sys} & : & 0 \\ \text{int}\,V^{mat} & : & 0 \\ \text{int}\,V^{met} & : & \vec{F}^{inc} \end{array}\right\} \begin{aligned} = & -\Big[\overset{\leftrightarrow}{G}{}^{JF}_{env}(\vec{r},\vec{r}\,')*\hat{e}_l \lim_{\vec{r}\,''\to\vec{r}\,'}\oint_{C(\vec{r}\,'')}\vec{H}^{inc}(\vec{r}\,'')\cdot d\vec{l}\,''\Big]_{L^{met}} \\ & +\Big[\overset{\leftrightarrow}{G}{}^{MF}_{env}(\vec{r},\vec{r}\,')*\hat{e}_l \lim_{\vec{r}\,''\to\vec{r}\,'}\oint_{C(\vec{r}\,'')}\vec{E}^{inc}(\vec{r}\,'')\cdot d\vec{l}\,''\Big]_{L^{met}} \\ & +\Big[\overset{\leftrightarrow}{G}{}^{JF}_{env}*\big(\hat{n}_{S^{met}_{-}}\times\vec{H}^{inc}\big)+\overset{\leftrightarrow}{G}{}^{MF}_{env}*\big(\vec{E}^{inc}\times\hat{n}_{S^{met}_{-}}\big)\Big]_{S^{met}_{+}} \\ & +\Big[\overset{\leftrightarrow}{G}{}^{JF}_{env}*\big(\hat{n}_{S^{met}_{+}}\times\vec{H}^{inc}\big)+\overset{\leftrightarrow}{G}{}^{MF}_{env}*\big(\vec{E}^{inc}\times\hat{n}_{S^{met}_{+}}\big)\Big]_{S^{met}_{-}} \\ & +\Big[\overset{\leftrightarrow}{G}{}^{JF}_{env}*\big(\hat{n}_{V^{met}_{-}}\times\vec{H}^{inc}\big)+\overset{\leftrightarrow}{G}{}^{MF}_{env}*\big(\vec{E}^{inc}\times\hat{n}_{V^{met}_{-}}\big)\Big]_{\partial V^{met}_{-}} \\ = & \Big[\overset{\leftrightarrow}{G}{}^{JF}_{env}*\big(\hat{n}_{V^{met}_{-}}\times\vec{H}^{inc}\big)+\overset{\leftrightarrow}{G}{}^{MF}_{env}*\big(\vec{E}^{inc}\times\hat{n}_{V^{met}_{-}}\big)\Big]_{\partial V^{met}} \end{aligned} \tag{41.2}$$

where the second equality is due to that the incident sources don't distribute on $L^{met}$, $S^{met}$, and $\partial V^{met}$; the mathematical formulation of the GHP corresponding to metallic scattering field is as follows:

$$\left.\begin{array}{lll}\text{ext}\,D^{sys} & : & \vec{F}^{sca}_{met}\\ \text{int}V^{mat} & : & \vec{F}^{sca}_{met}\\ \text{int}V^{met} & : & 0\end{array}\right\}=\begin{array}{l}\left[\overleftrightarrow{G}^{JF}_{env}(\vec{r},\vec{r}')*\hat{e}_l\lim\limits_{\vec{r}''\to\vec{r}'}\oint_{C(\vec{r}'')}\vec{H}^{sca}_{met}(\vec{r}'')\cdot d\vec{l}''\right]_{L^{met}}\\ -\left[\overleftrightarrow{G}^{MF}_{env}(\vec{r},\vec{r}')*\hat{e}_l\lim\limits_{\vec{r}''\to\vec{r}'}\oint_{C(\vec{r}'')}\vec{E}^{sca}_{met}(\vec{r}'')\cdot d\vec{l}''\right]_{L^{met}}\\ +\left[\overleftrightarrow{G}^{JF}_{env}*\left(\hat{n}_{S^{met}_+}\times\vec{H}^{sca}_{met}\right)+\overleftrightarrow{G}^{MF}_{env}*\left(\vec{E}^{sca}_{met}\times\hat{n}_{S^{met}_+}\right)\right]_{S^{met}_+}\\ +\left[\overleftrightarrow{G}^{JF}_{env}*\left(\hat{n}_{S^{met}_-}\times\vec{H}^{sca}_{met}\right)+\overleftrightarrow{G}^{MF}_{env}*\left(\vec{E}^{sca}_{met}\times\hat{n}_{S^{met}_{\cap-}}\right)\right]_{S^{met}_-}\\ +\left[\overleftrightarrow{G}^{JF}_{env}*\left(\hat{n}_{V^{met}_+}\times\vec{H}^{sca}_{met}\right)+\overleftrightarrow{G}^{MF}_{env}*\left(\vec{E}^{sca}_{met}\times\hat{n}_{V^{met}_+}\right)\right]_{\partial V^{met}_+}\end{array}$$

$$=\begin{array}{l}\left[\overleftrightarrow{G}^{JF}_{env}*\vec{J}^{SL}\right]_{L^{met}}+\left[\overleftrightarrow{G}^{JF}_{env}*\vec{J}^{SS}\right]_{S^{met}}\\ +\left[\overleftrightarrow{G}^{JF}_{env}*\left(\hat{n}_{V^{met}_+}\times\vec{H}^{sca}_{met}\right)+\overleftrightarrow{G}^{MF}_{env}*\left(\vec{E}^{sca}_{met}\times\hat{n}_{V^{met}_+}\right)\right]_{\partial V^{met}_+}\end{array}\tag{42.3}$$

where the second equality is due to that there doesn't exist metal-based scattering magnetic current on $L^{met}$ and $S^{met}$; the mathematical formulation of the GHP corresponding to material scattering field is as follows:

$$\left.\begin{array}{lll}\text{ext}\,D^{sys} & : & 0\\ \text{int}V^{mat} & : & 0\\ \text{int}V^{met} & : & \vec{F}^{sca}_{mat}\end{array}\right\}=\begin{array}{l}-\left[\overleftrightarrow{G}^{JF}_{env}(\vec{r},\vec{r}')*\hat{e}_l\lim\limits_{\vec{r}''\to\vec{r}'}\oint_{C(\vec{r}'')}\vec{H}^{sca}_{mat}(\vec{r}'')\cdot d\vec{l}''\right]_{L^{met}}\\ +\left[\overleftrightarrow{G}^{MF}_{env}(\vec{r},\vec{r}')*\hat{e}_l\lim\limits_{\vec{r}''\to\vec{r}'}\oint_{C(\vec{r}'')}\vec{E}^{sca}_{mat}(\vec{r}'')\cdot d\vec{l}''\right]_{L^{met}}\\ +\left[\overleftrightarrow{G}^{JF}_{env}*\left(\hat{n}_{S^{met}_-}\times\vec{H}^{sca}_{mat}\right)+\overleftrightarrow{G}^{MF}_{env}*\left(\vec{E}^{sca}_{mat}\times\hat{n}_{S^{met}_-}\right)\right]_{S^{met}_+}\\ +\left[\overleftrightarrow{G}^{JF}_{env}*\left(\hat{n}_{S^{met}_+}\times\vec{H}^{sca}_{mat}\right)+\overleftrightarrow{G}^{MF}_{env}*\left(\vec{E}^{sca}_{mat}\times\hat{n}_{S^{met}_+}\right)\right]_{S^{met}_-}\\ +\left[\overleftrightarrow{G}^{JF}_{env}*\left(\hat{n}_{V^{met}_-}\times\vec{H}^{sca}_{mat}\right)+\overleftrightarrow{G}^{MF}_{env}*\left(\vec{E}^{sca}_{mat}\times\hat{n}_{V^{met}_-}\right)\right]_{\partial V^{met}_-}\end{array}$$

$$=\left[\overleftrightarrow{G}^{JF}_{env}*\left(\hat{n}_{V^{met}_-}\times\vec{H}^{sca}_{mat}\right)+\overleftrightarrow{G}^{MF}_{env}*\left(\vec{E}^{sca}_{mat}\times\hat{n}_{V^{met}_-}\right)\right]_{\partial V^{met}}\tag{42.4}$$

where the second equality is due to that there doesn't exist material-based scattering current on $L^{met}$, $S^{met}$, and $\partial V^{met}$.

**3) The GHP and GET corresponding to the "Huygens' boundary" which is selected as system boundary: Topological additivity**

The summation of GHP (41.1) and (41.2) is the following incident field GHP (41'):

$$\left.\begin{array}{lll}\text{ext}\,D^{sys} & : & 0\\ \text{int}V^{mat} & : & \vec{F}^{inc}\\ \text{int}V^{met} & : & \vec{F}^{inc}\end{array}\right\}=\begin{array}{l}\left[\overleftrightarrow{G}^{JF}_{env}*\left(\hat{n}_{V^{mat}_-}\times\vec{H}^{inc}\right)+\overleftrightarrow{G}^{MF}_{env}*\left(\vec{E}^{inc}\times\hat{n}_{V^{mat}_-}\right)\right]_{\partial V^{mat}_0}\\ +\left[\overleftrightarrow{G}^{JF}_{env}*\left(\hat{n}_{V^{met}_-}\times\vec{H}^{inc}\right)+\overleftrightarrow{G}^{MF}_{env}*\left(\vec{E}^{inc}\times\hat{n}_{V^{met}_-}\right)\right]_{\partial V^{met}_0}\end{array}$$

$$=\left[\overleftrightarrow{G}^{JF}_{env}*\left(\hat{n}_{D^{sys}_-}\times\vec{H}^{inc}\right)+\overleftrightarrow{G}^{MF}_{env}*\left(\vec{E}^{inc}\times\hat{n}_{D^{sys}_-}\right)\right]_{\partial V^{mat}_0\cup\partial V^{met}_0}\tag{41'}$$

where the first equality is due to (4), (10), and that $\hat{n}_{V^{mat}_-}=-\hat{n}_{V^{met}_-}$ on $\partial V^{met}_{\cap}$, and that $\vec{F}^{inc}$ is continuous on $\partial V^{met}_{\cap}$; in the right-hand side of second equality, integral domain $\partial V^{mat}_0\cup\partial V^{met}_0$ is just the closed surface part of whole system boundary $\partial D^{sys}$, and $\hat{n}_{D^{sys}_-}$ is the inward normal vector of surface $\partial V^{mat}_0\cup\partial V^{met}_0$.

The summation of the material scattering field GHP (42.1)& (42.4) and the metallic scattering field GHP (42.2)&(42.3) is

$$\left.\begin{array}{lll}\text{ext}\,D^{sys} & : & \vec{F}^{sca}\\ \text{int}V^{mat} & : & 0\\ \text{int}V^{met} & : & 0\end{array}\right\}=\begin{array}{l}\left[\overleftrightarrow{G}^{JF}_{env}*\left(\hat{n}_{V^{mat}_+}\times\vec{H}^{sca}\right)+\overleftrightarrow{G}^{MF}_{env}*\left(\vec{E}^{sca}\times\hat{n}_{V^{mat}_+}\right)\right]_{\partial V^{mat}_{0-}}\\ +\left[\overleftrightarrow{G}^{JF}_{env}*\left(\hat{n}_{V^{met}_+}\times\vec{H}^{sca}\right)+\overleftrightarrow{G}^{MF}_{env}*\left(\vec{E}^{sca}\times\hat{n}_{V^{met}_+}\right)\right]_{\partial V^{met}_{0+}}\\ +\left[\overleftrightarrow{G}^{JF}_{env}*\vec{J}^{SS}_0\right]_{S^{met}_0}+\left[\overleftrightarrow{G}^{JF}_{env}*\vec{J}^{SL}_0\right]_{L^{met}_0}\end{array}$$

$$=\begin{array}{l}\left[\overleftrightarrow{G}^{JF}_{env}*\left(\hat{n}_{D^{sys}_+}\times\vec{H}^{sca}\right)+\overleftrightarrow{G}^{MF}_{env}*\left(\vec{E}^{sca}\times\hat{n}_{D^{sys}_+}\right)\right]_{\partial V^{mat}_{0-}\cup\partial V^{met}_{0+}}\\ +\left[\overleftrightarrow{G}^{JF}_{env}*\vec{J}^{SS}_0\right]_{S^{met}_0}+\left[\overleftrightarrow{G}^{JF}_{env}*\vec{J}^{SL}_0\right]_{L^{met}_0}\end{array}\tag{42'}$$

where $\hat{n}_{D^{sys}_+}$ is the outward normal vector of $\partial V^{mat}_0\cup\partial V^{met}_0$.

The above (41') and (42') are just the incident field GHP and the total scattering field GHP corresponding to "Huygens' boundary" $\partial D^{sys}$. Obviously, they satisfy GET and the topological additivity introduced in paper [6].

*B. Generalized Franz-Harrington formulation and artificial extinction theorem*

The summation of the incident field GHP (41') and scattering field GHP (42') gives the following GFHF of internal incident field and external scattering field.

$$\left.\begin{array}{lll}\text{ext}\,D^{sys} & : & -\vec{F}^{sca}\\ \text{int}V^{mat} & : & \vec{F}^{inc}\\ \text{int}V^{met} & : & \vec{F}^{inc}\end{array}\right\}=\begin{array}{l}\left[\overleftrightarrow{G}^{JF}_{env}*\vec{J}^{ES}_0+\overleftrightarrow{G}^{MF}_{env}*\vec{M}^{ES}_0\right]_{\partial V^{mat}_0}\\ -\left[\overleftrightarrow{G}^{JF}_{env}*\vec{J}^{SS}_0\right]_{S^{met}_0\cup\partial V^{met}_0}-\left[\overleftrightarrow{G}^{JF}_{env}*\vec{J}^{SL}_0\right]_{L^{met}_0}\end{array}.\tag{43}$$

Following the ideas of paper [6], the following *piecewise Green's functions* are proposed to derive so-called *artificial extinction theorem* corresponding to internal total field.

$$\tilde{\overleftrightarrow{G}}^{JF}_{sys}(\vec{r},\vec{r}')\triangleq\begin{cases}\overleftrightarrow{G}^{JF}_{mat}(\vec{r},\vec{r}') & ,\ \left(\vec{r}\in\text{cl}V^{mat}\ ,\vec{r}'\in\text{cl}V^{mat}\right)\\ 0 & ,\ \left(\vec{r}\in\text{ext}V^{mat}\ ,\vec{r}'\in\text{cl}V^{mat}\right)\\ \overleftrightarrow{G}^{JF}_{env}(\vec{r},\vec{r}') & ,\ \left(\vec{r}'\in\text{ext}V^{mat}\right)\end{cases}\tag{44.1}$$

$$\tilde{\overleftrightarrow{G}}^{MF}_{sys}(\vec{r},\vec{r}')\triangleq\begin{cases}\overleftrightarrow{G}^{MF}_{mat}(\vec{r},\vec{r}') & ,\ \left(\vec{r}\in\text{cl}V^{mat}\ ,\vec{r}'\in\text{cl}V^{mat}\right)\\ 0 & ,\ \left(\vec{r}\in\text{ext}V^{mat}\ ,\vec{r}'\in\text{cl}V^{mat}\right)\\ \overleftrightarrow{G}^{MF}_{env}(\vec{r},\vec{r}') & ,\ \left(\vec{r}'\in\text{ext}V^{mat}\right)\end{cases}\tag{44.2}$$

where $\overleftrightarrow{G}^{JF}_{mat}$ and $\overleftrightarrow{G}^{MF}_{mat}$ are the Green's functions corresponding to the material part of EM system. Based on the above (44) and the results given in paper [6], the following artificial extinction theorem for internal total field exists:

$$\left.\begin{array}{lll}\text{ext}\,D^{sys} & : & 0\\ \text{int}V^{mat} & : & \vec{F}^{tot}\\ \text{int}V^{met} & : & 0\end{array}\right\}=\begin{array}{l}\left[\tilde{\overleftrightarrow{G}}^{JF}_{sys}*\vec{J}^{EL}\right]_{L^{met}_{\cap}}+\left[\tilde{\overleftrightarrow{G}}^{JF}_{sys}*\vec{J}^{ES}_{\cap}\right]_{S^{met}_{\cap}\cup\partial V^{met}_{\cap}}\\ +\left[\tilde{\overleftrightarrow{G}}^{JF}_{sys}*\vec{J}^{ES}_0+\tilde{\overleftrightarrow{G}}^{MF}_{sys}*\vec{M}^{ES}_0\right]_{\partial V^{mat}_0}\end{array}.\tag{45}$$

In fact, the above (45) can be equivalently rewritten as follows:

$$\left.\begin{array}{lll}\text{ext}\,D^{sys} & : & 0\\ \text{int}V^{mat} & : & \vec{F}^{tot}\\ \text{int}V^{met} & : & \vec{F}^{tot}\end{array}\right\}=\begin{array}{l}\left[\tilde{\overleftrightarrow{G}}^{JF}_{sys}*\vec{J}^{SL}_{\cap}\right]_{L^{met}_{\cap}}+\left[\tilde{\overleftrightarrow{G}}^{JF}_{sys}*\vec{J}^{SS}_{\cap}\right]_{S^{met}_{\cap}\cup\partial V^{met}_{\cap}}\\ +\left[\tilde{\overleftrightarrow{G}}^{JF}_{sys}*\vec{J}^{ES}_0+\tilde{\overleftrightarrow{G}}^{MF}_{sys}*\vec{M}^{ES}_0\right]_{\partial V^{mat}_0}\end{array}\tag{45'}$$

because of (36.1), (38.1), and that $\vec{F}^{tot} \equiv 0$ on $\mathrm{int}V^{met}$.

Following the ideas of paper [6], the following *piecewise delta Green's functions* are proposed:

$$\Delta\tilde{\vec{G}}_{sys}^{JF}(\vec{r},\vec{r}') \triangleq \tilde{\vec{G}}_{sys}^{JF}(\vec{r},\vec{r}') - \vec{G}_{env}^{JF}(\vec{r},\vec{r}') = \begin{cases} \vec{G}_{mat}^{JF}(\vec{r},\vec{r}') - \vec{G}_{env}^{JF}(\vec{r},\vec{r}') &, (\vec{r} \in \mathrm{cl}V^{mat}, \vec{r}' \in \mathrm{cl}V^{mat}) \\ -\vec{G}_{env}^{JF}(\vec{r},\vec{r}') &, (\vec{r} \in \mathrm{ext}V^{mat}, \vec{r}' \in \mathrm{cl}V^{mat}) \\ 0 &, (\vec{r}' \in \mathrm{ext}V^{mat}) \end{cases} \tag{46.1}$$

$$\Delta\tilde{\vec{G}}_{sys}^{MF}(\vec{r},\vec{r}') \triangleq \tilde{\vec{G}}_{sys}^{MF}(\vec{r},\vec{r}') - \vec{G}_{env}^{MF}(\vec{r},\vec{r}') = \begin{cases} \vec{G}_{mat}^{MF}(\vec{r},\vec{r}') - \vec{G}_{env}^{MF}(\vec{r},\vec{r}') &, (\vec{r} \in \mathrm{cl}V^{mat}, \vec{r}' \in \mathrm{cl}V^{mat}) \\ -\vec{G}_{env}^{MF}(\vec{r},\vec{r}') &, (\vec{r} \in \mathrm{ext}V^{mat}, \vec{r}' \in \mathrm{cl}V^{mat}) \\ 0 &, (\vec{r}' \in \mathrm{ext}V^{mat}) \end{cases} \tag{46.2}$$

and then the following GFHF (47) of scattering field can be obtained:

$$\left.\begin{array}{ll} \mathrm{ext}D^{sys} : & \vec{F}^{sca} \\ \mathrm{int}V^{mat} : & \vec{F}^{sca} \\ \mathrm{int}V^{met} : & \vec{F}^{sca} \end{array}\right\} = \left[\tilde{\vec{G}}_{sys}^{JF} * \vec{J}^{SL}\right]_{L^{met}} + \left[\tilde{\vec{G}}_{sys}^{JF} * \vec{J}^{SS}\right]_{S^{met}\cup\partial V^{met}} + \left[\Delta\tilde{\vec{G}}_{sys}^{JF} * \vec{J}_0^{ES} + \Delta\tilde{\vec{G}}_{sys}^{MF} * \vec{M}_0^{ES}\right]_{\partial V_0^{mat}} \tag{47}$$

based on (43) and (45').

## C. Summary

In summary, above GHP, GET, and GFHF of metal-material combined system are formally identical to the inhomogeneous anisotropic lossy material system given in Part I, and the former satisfies the same topological additivity as the latter, i.e.,

$$\begin{aligned} &\text{Scattering field GHP/GET of metal-material combined system} \\ =\ &\text{Scattering field GHP/GET of metallic subsystem} \\ +\ &\text{Scattering field GHP/GET of material subsystem} \\ =\ &\textstyle\sum_{\xi} \text{Scattering field GHP/GET of metallic line } L_{\xi}^{met} \\ +\ &\textstyle\sum_{\zeta} \text{Scattering field GHP/GET of metallic surface } S_{\zeta}^{met} \\ +\ &\textstyle\sum_{\upsilon} \text{Scattering field GHP/GET of metallic body } V_{\upsilon}^{met} \\ +\ &\textstyle\sum_{\nu} \text{Scattering field GHP/GET of material body } V_{\nu}^{mat} \end{aligned} \tag{48.1}$$

and

$$\begin{aligned} &\text{Incident field GHP/GET of metal-material combined system} \\ =\ &\text{Incident field GHP/GET of metallic subsystem} \\ +\ &\text{Incident field GHP/GET of material subsystem} \\ =\ &\textstyle\sum_{\xi} \text{Incident field GHP/GET of metallic line } L_{\xi}^{met} \\ +\ &\textstyle\sum_{\zeta} \text{Incident field GHP/GET of metallic surface } S_{\zeta}^{met} \\ +\ &\textstyle\sum_{\upsilon} \text{Incident field GHP/GET of metallic body } V_{\upsilon}^{met} \\ +\ &\textstyle\sum_{\nu} \text{Incident field GHP/GET of material body } V_{\nu}^{mat} \end{aligned} \tag{48.2}$$

and

$$\begin{aligned} &\text{The GFHF of metal-material combined system} \\ =\ &\text{The GFHF of metallic subsystem} \\ +\ &\text{The GFHF of material subsystem} \\ =\ &\textstyle\sum_{\xi} \text{The GFHF of metallic line } L_{\xi}^{met} \\ +\ &\textstyle\sum_{\zeta} \text{The GFHF of metallic surface } S_{\zeta}^{met} \\ +\ &\textstyle\sum_{\upsilon} \text{The GFHF of metallic body } V_{\upsilon}^{met} \\ +\ &\textstyle\sum_{\nu} \text{The GFHF of material body } V_{\nu}^{mat} \end{aligned} \ . \tag{48.3}$$

As the typical engineering applications, the above GFHF is applied to construct the CM of metal-material combined system in the following Secs. VI and VII. In the following Secs. VI and VII, the environment is restricted to being VACUUM, and then the various environment Green's functions become free-space Green's functions, and the mathematical formulations of these free-space Green's functions can be found in [8].

# VI. Application of GFHF: To Construct Harrington's CM of Metal-Material Combined System

For metallic system, Harrington *et al.* [11] developed a mathematical scheme to construct CM by using SEFIE-MoM (surface electric field integral equation based method of moments). For isotropic material system, Harrington *et al.* constructed some kinds of CM by using VIE-MoM (volume integral equation based MoM) [12] and SIE-MoM (surface integral equation based MoM, also known as PMCHWT-based MoM) [5]. The physical essence of *Harrington's CM* is to construct a series of orthogonal modes which have ability to orthogonalize objective EM power, for example:

• For metallic system, Harrington's SEFIE-based CM [11] orthogonalizes the following objective power:

$$(1/2)\left\langle \vec{J}^{SL}, \vec{E}^{inc} \right\rangle_{L^{met}} + (1/2)\left\langle \vec{J}^{SS}, \vec{E}^{inc} \right\rangle_{S^{met}\cup\partial V^{met}} \tag{49}$$

where the inner product is defined as $< \vec{f}, \vec{g} >_{\Omega} \triangleq \int_{\Omega} \vec{f}^* \cdot \vec{g} \, d\Omega$.

• For homogeneous or inhomogeneous isotropic material system, Harrington's VIE-based CM [12] orthogonalizes the following objective power:

$$(1/2)\left\langle \vec{J}^{SV}, \vec{E}^{inc} \right\rangle_{V^{mat}} + (1/2)\left\langle \vec{M}^{SV}, \vec{H}^{inc} \right\rangle_{V^{mat}} . \tag{50}$$

• For homogeneous isotropic material system, Harrington's PMCHWT-based CM [5] orthogonalizes the following power:

$$-(1/2)\left\langle \vec{J}^{ES}, \vec{E}^{inc} \right\rangle_{\partial V^{mat}} - (1/2)\left\langle \vec{M}^{ES}, \vec{H}^{inc} \right\rangle_{\partial V^{mat}} \tag{51}$$

where the minus signs originate from that the equivalent surface currents in [5] are $\{-\vec{J}^{ES}, -\vec{M}^{ES}\}$.

Recently, [13] proved that the objective powers orthogonalized by VIE-based CM and PMCHWT-based CM are identical to each other, i.e.,

$$\frac{1}{2}\left\langle \vec{J}^{SV}, \vec{E}^{inc} \right\rangle_{V^{mat}} + \frac{1}{2}\left\langle \vec{M}^{SV}, \vec{H}^{inc} \right\rangle_{V^{mat}} = -\frac{1}{2}\left\langle \vec{J}^{ES}, \vec{E}^{inc} \right\rangle_{\partial V^{mat}} - \frac{1}{2}\left\langle \vec{M}^{ES}, \vec{H}^{inc} \right\rangle_{\partial V^{mat}} \tag{52}$$

when material system is homogeneous isotropic. By directly combining Harrington's SEFIE-based and PMCHWT-based schemes, [14] constructed the SEFIE-PMCHWT-based CM of the metallic body which was completely coated by a homogeneous isotropic lossless dielectric medium.

In this section, Harrington's CM will be further generalized to the metal-material combined system whose metallic part can be line or surface or body and material part can be inhomogeneous anisotropic lossy material body. The reason to call the CM constructed below as "Harrington's CM" is that the CM orthogonalizes the following objective power operator:

$$P_{met-mat\ sys}^{Harrington} = \frac{1}{2}\left\langle \vec{J}^{SL} \oplus \vec{J}^{SS}, \vec{E}^{inc} \right\rangle_{L^{met} \cup S^{met} \cup \partial V^{met}} + \frac{1}{2}\left\langle \vec{J}^{SV}, \vec{E}^{inc} \right\rangle_{V^{mat}} + \frac{1}{2}\left\langle \vec{M}^{SV}, \vec{H}^{inc} \right\rangle_{V^{mat}} \tag{53}$$

by following Harrington's ideas in [11] and [12]. The subscript "$met-mat\ sys$" in $P_{met-mat\ sys}^{Harrington}$ is to emphasize that the power corresponds to metal-material combined system. The symbol "$\oplus$" in $< \vec{J}^{SL} \oplus \vec{J}^{SS}, \vec{E}^{inc} >_{L^{met} \cup S^{met} \cup \partial V^{met}}$ is defined as follows:

$$\left\langle \vec{J}^{SL} \oplus \vec{J}^{SS}, \vec{E}^{inc} \right\rangle_{L^{met} \cup S^{met} \cup \partial V^{met}} \triangleq \left\langle \vec{J}^{SL}, \vec{E}^{inc} \right\rangle_{L^{met}} + \left\langle \vec{J}^{SS}, \vec{E}^{inc} \right\rangle_{S^{met} \cup \partial V^{met}} \tag{54}$$

and the reason to utilize "$\oplus$" instead of "$+$" is that the dimensions of line current $\vec{J}^{SL}$ and surface current $\vec{J}^{SS}$ are different from each other.

### *A. Power characteristic of operator (53)*

The power operator $P_{met-mat\ sys}^{Harrington}$ in (53) can be equivalently rewritten as follows:

$$\begin{aligned} P_{met-mat\ sys}^{Harrington} &= \frac{1}{2}\left\langle \vec{J}^{SL} \oplus \vec{J}^{SS}, \vec{E}^{inc} \right\rangle_{L^{met} \cup S^{met} \cup \partial V^{met}} + \frac{1}{2}\left\langle \vec{J}^{SV}, \vec{E}^{inc} \right\rangle_{V^{mat}} + \frac{1}{2}\left\langle \vec{M}^{SV}, \vec{H}^{inc} \right\rangle_{V^{mat}} \\ &\quad + (1/2)\left\langle \vec{H}^{inc}, \vec{M}^{SV} \right\rangle_{V^{mat}} - (1/2)\left\langle \vec{M}^{SV}, \vec{H}^{inc} \right\rangle_{V^{mat}}^{*} \\ &= \frac{1}{2}\left\langle \vec{J}^{SL} \oplus \vec{J}^{SS}, \vec{E}^{inc} \right\rangle_{L^{met} \cup S^{met} \cup \partial V^{met}} + \frac{1}{2}\left\langle \vec{J}^{SV}, \vec{E}^{inc} \right\rangle_{V^{mat}} + \frac{1}{2}\left\langle \vec{H}^{inc}, \vec{M}^{SV} \right\rangle_{V^{mat}} \\ &\quad + j\,\mathrm{Im}\left\{ \left\langle \vec{M}^{SV}, \vec{H}^{inc} \right\rangle_{V^{mat}} \right\} \\ &= -\frac{1}{2}\left\langle \vec{J}^{SL} \oplus \vec{J}^{SS}, \vec{E}^{sca} \right\rangle_{L^{met} \cup S^{met} \cup \partial V^{met}} - \frac{1}{2}\left\langle \vec{J}^{SV}, \vec{E}^{sca} \right\rangle_{V^{mat}} - \frac{1}{2}\left\langle \vec{H}^{sca}, \vec{M}^{SV} \right\rangle_{V^{mat}} \\ &\quad + (1/2)\left\langle \vec{J}^{SV}, \vec{E}^{tot} \right\rangle_{V^{mat}} + (1/2)\left\langle \vec{H}^{tot}, \vec{M}^{SV} \right\rangle_{V^{mat}} \\ &\quad + j\,\mathrm{Im}\left\{ \left\langle \vec{M}^{SV}, \vec{H}^{inc} \right\rangle_{V^{mat}} \right\} \end{aligned} \tag{53'}$$

In (53'), the first equality is due to that $< \vec{H}^{inc}, \vec{M}^{SV} >_{V^{mat}} = < \vec{M}^{SV}, \vec{H}^{inc} >_{V^{mat}}^{*}$ ; the second equality is because of that $C - C^{*} = j2\,\mathrm{Im}\{C\}$ for any complex scalar $C$; the third equality is based on the tangential electric field boundary condition on metallic boundary and that $\vec{F}^{inc} = \vec{F}^{tot} - \vec{F}^{sca}$ on $V^{mat}$.

Based on complex Poynting's theorem [4], constitutive relationship, and that $\Delta\vec{\mu}$ is real symmetrical, the following relations can be derived:

$$-\frac{1}{2}\left\langle \vec{J}^{SL} \oplus \vec{J}^{SS}, \vec{E}^{sca} \right\rangle_{L^{met} \cup S^{met} \cup \partial V^{met}} - \frac{1}{2}\left\langle \vec{J}^{SV}, \vec{E}^{sca} \right\rangle_{V^{mat}} - \frac{1}{2}\left\langle \vec{H}^{sca}, \vec{M}^{SV} \right\rangle_{V^{mat}} = P_{met-mat\ sys}^{sca,rad} + j\, P_{met-mat\ sys}^{sca,sto,field} \tag{55.1}$$

$$(1/2)\left\langle \vec{J}^{SV}, \vec{E}^{tot} \right\rangle_{V^{mat}} + (1/2)\left\langle \vec{H}^{tot}, \vec{M}^{SV} \right\rangle_{V^{mat}} = P_{met-mat\ sys}^{tot,loss,mat} + j\, P_{met-mat\ sys}^{tot,sto,mat} \tag{55.2}$$

$$\mathrm{Im}\left\{ \left\langle \vec{M}^{SV}, \vec{H}^{inc} \right\rangle_{V^{mat}} \right\} = -\omega\left[ \left\langle \Delta\vec{\mu} \cdot \vec{H}^{inc}, \vec{H}^{inc} \right\rangle_{V^{mat}} + \mathrm{Re}\left\{ \left\langle \Delta\vec{\mu} \cdot \vec{H}^{sca}, \vec{H}^{inc} \right\rangle_{V^{mat}} \right\} \right] \tag{55.3}$$

where

$$P_{met-mat\ sys}^{sca,rad} = (1/2) \oiint_{S_\infty} \left[ \vec{E}^{sca} \times \left( \vec{H}^{sca} \right)^{*} \right] \cdot d\vec{S} \tag{56.1}$$

$$P_{met-mat\ sys}^{sca,sto,field} = 2\omega\left( W_{met-mat\ sys;m}^{sca,sto,field} - W_{met-mat\ sys;e}^{sca,sto,field} \right) \tag{56.2}$$

$$P_{met-mat\ sys}^{tot,loss,mat} = (1/2)\left\langle \vec{\sigma} \cdot \vec{E}^{tot}, \vec{E}^{tot} \right\rangle_{V^{mat}} \tag{56.3}$$

$$P_{met-mat\ sys}^{tot,sto,mat} = 2\omega\left( W_{met-mat\ sys;m}^{tot,sto,mat} - W_{met-mat\ sys;e}^{tot,sto,mat} \right) \tag{56.4}$$

in which $S_\infty$ is a spherical surface at infinity, and

$$W_{met-mat\ sys;m}^{sca,sto,field} = (1/4)\left\langle \vec{H}^{sca}, \mu_0 \vec{H}^{sca} \right\rangle_{\mathbb{R}^3} \tag{57.1}$$

$$W_{met-mat\ sys;e}^{sca,sto,field} = (1/4)\left\langle \varepsilon_0 \vec{E}^{sca}, \vec{E}^{sca} \right\rangle_{\mathbb{R}^3} \tag{57.2}$$

$$W_{met-mat\ sys;m}^{tot,sto,mat} = (1/4)\left\langle \vec{H}^{tot}, \Delta\vec{\mu} \cdot \vec{H}^{tot} \right\rangle_{V^{mat}} \tag{57.3}$$

$$W_{met-mat\ sys;e}^{tot,sto,mat} = (1/4)\left\langle \Delta\vec{\varepsilon} \cdot \vec{E}^{tot}, \vec{E}^{tot} \right\rangle_{V^{mat}} . \tag{57.4}$$

In (56.1), the superscript "$sca,rad$" means that $P_{met-mat\ sys}^{sca,rad}$ is the radiated power carried by scattering field; in (56.2), the superscript "$sca,sto,field$" and subscript "$m/e$" mean that $W_{met-mat\ sys;m/e}^{sca,sto,field}$ is the magnetic/electric energy stored in scattering field; in (56.3), the superscript "$tot,loss,mat$" means that $P_{met-mat\ sys}^{tot,loss,mat}$ is the lossy power due to the interaction between total electric field and material; in (56.4), the superscript "$tot,sto,mat$" and subscript "$m/e$" mean that $W_{met-mat\ sys;m/e}^{tot,sto,mat}$ is the magnetization/polarization energy due to the interaction between total magnetic/electric field and material. In (57.3) and (57.4), $\Delta\vec{\mu} \triangleq \vec{\mu} - \vec{I}\mu_0$ and $\Delta\vec{\varepsilon} \triangleq \vec{\varepsilon} - \vec{I}\varepsilon_0$ .

Inserting (55) into the last equality of (53'), the power characteristic of operator $P_{met-mat\ sys}^{Harrington}$ is exhibited as below:

$$\begin{aligned} P_{met-mat\ sys}^{Harrington} = P_{met-mat\ sys}^{sca,rad} + P_{met-mat\ sys}^{tot,loss,mat} + j\left( P_{met-mat\ sys}^{sca,sto,field} + P_{met-mat\ sys}^{tot,sto,mat} \right) \\ - j\omega\left[ \left\langle \Delta\vec{\mu} \cdot \vec{H}^{inc}, \vec{H}^{inc} \right\rangle_{V^{mat}} + \mathrm{Re}\left\{ \left\langle \Delta\vec{\mu} \cdot \vec{H}^{sca}, \vec{H}^{inc} \right\rangle_{V^{mat}} \right\} \right] \end{aligned} . \tag{58}$$

### *B. Line-surface formulation of operator (53)*

Based on the same process as deriving (64.1) and (64.2) in paper [6], the following relations corresponding to the material body shown in Fig. 1 can be derived:

$$\left\langle \vec{J}^{EL} \oplus \vec{J}^{ES}, \vec{E}^{inc} \right\rangle_{\partial V^{mat}} = -j\omega\left\langle \vec{H}^{tot}, \mu_0 \vec{H}^{inc} \right\rangle_{V^{mat}} + j\omega\left\langle \varepsilon_0 \vec{E}^{tot}, \vec{E}^{inc} \right\rangle_{V^{mat}} - \left\langle \vec{J}^{SV}, \vec{E}^{inc} \right\rangle_{V^{mat}} \tag{59.1}$$

$$\left\langle \vec{M}^{EL} \oplus \vec{M}^{ES}, \vec{H}^{inc} \right\rangle_{\partial V^{mat}} = j\omega\left\langle \vec{H}^{tot}, \mu_0 \vec{H}^{inc} \right\rangle_{V^{mat}} - j\omega\left\langle \varepsilon_0 \vec{E}^{tot}, \vec{E}^{inc} \right\rangle_{V^{mat}} - \left\langle \vec{M}^{SV}, \vec{H}^{inc} \right\rangle_{V^{mat}} . \tag{59.2}$$

The summation of (59.1) and (59.2) gives that

$$\begin{aligned}&\left\langle \vec{J}^{SV}, \vec{E}^{inc}\right\rangle_{V^{mat}} + \left\langle \vec{M}^{SV}, \vec{H}^{inc}\right\rangle_{V^{mat}}\\ &= -\left\langle \vec{J}^{EL}\oplus\vec{J}^{ES}, \vec{E}^{inc}\right\rangle_{\partial V^{mat}} - \left\langle \vec{M}^{EL}\oplus \vec{M}^{ES}, \vec{H}^{inc}\right\rangle_{\partial V^{mat}}\\ &= -\left\langle \vec{J}^{SL}_{\cap}\oplus\left(\vec{J}^{SS}_{\cap}+\vec{J}^{ES}_0\right), \vec{E}^{inc}\right\rangle_{L^{met}_\cap \cup S^{met}_\cap\cup \partial V^{met}_\cap \cup \partial V^{mat}_0} - \left\langle \vec M_0^{ES}, \vec H^{inc}\right\rangle_{\partial V_0^{mat}}\end{aligned} \tag{60}$$

where the second equality is based on (14), (31), (36), and (38).

Inserting (60) into (53), it is obtained that

$$\begin{aligned}&P^{Harrington}_{met-mat\ sys}\\ &= -(1/2)\left\langle \vec J_0^{SL}\oplus \vec J_0^{SS}, \vec E^{sca}\right\rangle_{L_0^{met}\cup S_0^{met}\cup \partial V_0^{met}}\\ &\quad +(1/2)\left\langle \vec J_\cap^{SL}\oplus \vec J_\cap^{SS}, \vec E^{inc}\right\rangle_{L_\cap^{met}\cup S_\cap^{met}\cup \partial V_\cap^{met}}\\ &\quad -(1/2)\left\langle \vec J_\cap^{SL}\oplus \left(\vec J_\cap^{SS}+\vec J_0^{ES}\right), \vec E^{inc}\right\rangle_{L_\cap^{met}\cup S_\cap^{met}\cup \partial V_\cap^{met}\cup\partial V_0^{mat}} -(1/2)\left\langle \vec M_0^{ES}, \vec H^{inc}\right\rangle_{\partial V_0^{mat}}\\ &= -(1/2)\left\langle \vec J_0^{SL}\oplus \vec J_0^{SS}, \vec E^{sca}\right\rangle_{L_0^{met}\cup S_0^{met}\cup \partial V_0^{met}}\\ &\quad -(1/2)\left\langle \vec J_0^{ES}, \vec E^{inc}\right\rangle_{\partial V_0^{mat}} -(1/2)\left\langle \vec M_0^{ES}, \vec H^{inc}\right\rangle_{\partial V_0^{mat}}\end{aligned} \tag{53''}$$

where the first equality is based on (15)-(16) and electric field tangential boundary condition; the second equality is obvious. By utilizing the GFHF given in (43), the line-surface formulation of power operator $P^{Harrington}_{met-mat\ sys}$ can be expressed as follows:

$$\begin{aligned}&P^{Harrington}_{met-mat\ sys}\\ &= (1/2)\left\langle \vec J_0^{SL}\oplus \vec J_0^{SS}\ ,\ j\omega\mu_0\mathcal{L}_0\left(\vec J_0^{SL}\oplus \vec J_0^{SS}-\vec J_0^{ES}\right)-\mathcal{K}_0\left(\vec M_0^{ES}\right)\right\rangle_{L_0^{met}\cup S_0^{met}\cup\partial V_0^{met}}\\ &\quad -(1/2)\left\langle \vec J_0^{ES}\ ,\ j\omega\mu_0\mathcal{L}_0\left(\vec J_0^{SL}\oplus \vec J_0^{SS}-\vec J_0^{ES}\right)-\mathcal{K}_0\left(\vec M_0^{ES}\right)\right\rangle_{\partial V_{0-}^{mat}}\\ &\quad -(1/2)\left\langle \vec M_0^{ES}\ ,\ -j\omega\varepsilon_0\mathcal{L}_0\left(\vec M_0^{ES}\right)-\mathcal{K}_0\left(\vec J_0^{SL}\oplus \vec J_0^{SS}-\vec J_0^{ES}\right)\right\rangle_{\partial V_{0-}^{mat}}\end{aligned} \tag{53'''}$$

where [5], [8]

$$\mathcal{L}_0\left(\vec X\right) \triangleq \left(1+\frac{1}{k_0^2}\nabla\nabla\cdot\right)\int_\Pi G_0\left(\vec r,\vec r'\right)\vec X\left(\vec r'\right)d\Pi' \tag{61.1}$$

$$\mathcal{K}_0\left(\vec X\right) \triangleq \nabla\times\int_\Pi G_0\left(\vec r,\vec r'\right)\vec X\left(\vec r'\right)d\Pi' \tag{61.2}$$

and $G_0(\vec r,\vec r') = e^{-jk_0|\vec r-\vec r'|}/4\pi|\vec r-\vec r'|$, and $k_0=\omega\sqrt{\mu_0\varepsilon_0}$. The subscript "$-$" used in integral domain $\partial V_{0-}^{mat}$ is to emphasize that the integral is done on the internal surface of boundary $\partial V_0^{mat}$, because $\mathcal{K}_0\left(\vec C_0^{ES}\right)$ is discontinuous on the two sides of $\partial V_0^{mat}$. The reason to call (53''') as line-surface formulation is that all arguments in this formulation are line or surface currents.

## C. *Discretization of operator (53''')*

In this subsection, the operator (53''') is transformed from current space to expansion vector space at first, and then the equivalent electric and magnetic currents are related to each other in expansion vector space [13].

**From current space to expansion vector space**

To discretize the operator (53'''), the currents $\vec J_{0/\cap}^{SL}$, $\vec J_{0/\cap}^{SS}$, $\vec J_0^{ES}$, and $\vec M_0^{ES}$ are expanded in terms of proper basis functions as

$$\vec J_{0/\cap}^{SL}\left(\vec r\right) = \sum_{\xi=1}^{\Xi^{\vec J_{0/\cap}^{SL}}} a_\xi^{\vec J_{0/\cap}^{SL}}\vec b_\xi^{\vec J_{0/\cap}^{SL}}\left(\vec r\right) = \bar B^{\vec J_{0/\cap}^{SL}}\cdot \bar a^{\vec J_{0/\cap}^{SL}}\ ,\quad \left(\vec r\in L_{0/\cap}^{met}\right) \tag{62.1}$$

$$\vec J_{0/\cap}^{SS}\left(\vec r\right) = \sum_{\xi=1}^{\Xi^{\vec J_{0/\cap}^{SS}}} a_\xi^{\vec J_{0/\cap}^{SS}}\vec b_\xi^{\vec J_{0/\cap}^{SS}}\left(\vec r\right) = \bar B^{\vec J_{0/\cap}^{SS}}\cdot \bar a^{\vec J_{0/\cap}^{SS}}\ ,\quad \left(\vec r\in S_{0/\cap}^{met}\cup\partial V_{0/\cap}^{met}\right) \tag{62.2}$$

$$\vec C_0^{ES}\left(\vec r\right) = \sum_{\xi=1}^{\Xi^{\vec C_0^{ES}}} a_\xi^{\vec C_0^{ES}}\vec b_\xi^{\vec C_0^{ES}}\left(\vec r\right) = \bar B^{\vec C_0^{ES}}\cdot \bar a^{\vec C_0^{ES}}\ ,\quad \left(\vec r\in \partial V_0^{mat}\right) \tag{62.3}$$

where $C=J,M$, and

$$\bar B^X = \left[\vec b_1^X\ ,\ \vec b_2^X\ ,\ \cdots\ ,\ \vec b_{\Xi^X}^X\right] \tag{63.1}$$

$$\bar a^X = \left[a_1^X\ ,\ a_2^X\ ,\ \cdots\ ,\ a_{\Xi^X}^X\right]^T \tag{63.2}$$

for any $X=\vec J_{0/\cap}^{SL},\vec J_{0/\cap}^{SS},\vec J_0^{ES},\vec M_0^{ES}$. The symbol "$\cdot$" represents the matrix multiplication, and the superscript "$T$" in (63.2) represents the transpose of matrix. In addition, it must be EMPHASIZED that: $\vec b_\xi^{\vec M_0^{ES}}\left(\vec r\right)=0$ for any $\vec r\in\partial V_0^{mat}\cap S^{met}$, because of the tangential boundary condition of the total electric field on $S^{met}$; $\left[\hat n_{\rightarrow mat}\left(\vec r\right)\times \hat e_l\left(\vec r\right)\right]\cdot \vec b_\xi^{\vec M_0^{ES}}\left(\vec r\right)=0$ for any $\vec r\in \partial V_0^{mat}\cap L^{met}$, because of the tangential boundary condition of the total electric field on $L^{met}$.

Inserting (62) into (53'''), the objective power $P^{Harrington}_{met-mat\ sys}$ is discretized to the following matrix form:

$$\begin{aligned}P^{Harrington}_{met-mat\ sys} &= \left(\bar a_{met-mat\ sys}^{\{\vec J_0^{SL},\vec J_0^{SS},\vec J_0^{ES},\vec J_\cap^{SL},\vec J_\cap^{SS},\vec M_0^{ES}\}}\right)^H\\ &\quad\cdot \bar{\bar P}_{met-mat\ sys}^{\{\vec J_0^{SL},\vec J_0^{SS},\vec J_0^{ES},\vec J_\cap^{SL},\vec J_\cap^{SS},\vec M_0^{ES}\}}\cdot \bar a_{met-mat\ sys}^{\{\vec J_0^{SL},\vec J_0^{SS},\vec J_0^{ES},\vec J_\cap^{SL},\vec J_\cap^{SS},\vec M_0^{ES}\}}\end{aligned} \tag{64}$$

where superscript "$H$" represents transpose conjugate, and

$$\bar{\bar P}_{met-mat\ sys}^{\{\vec J_0^{SL},\vec J_0^{SS},\vec J_0^{ES},\vec J_\cap^{SL},\vec J_\cap^{SS},\vec M_0^{ES}\}} = \begin{bmatrix} \bar{\bar P}^{\vec J_0^{SL}\vec J_0^{SL}} & \bar{\bar P}^{\vec J_0^{SL}\vec J_0^{SS}} & \bar{\bar P}^{\vec J_0^{SL}\vec J_0^{ES}} & 0 & 0 & \bar{\bar P}^{\vec J_0^{SL}\vec M_0^{ES}}\\ \bar{\bar P}^{\vec J_0^{SS}\vec J_0^{SL}} & \bar{\bar P}^{\vec J_0^{SS}\vec J_0^{SS}} & \bar{\bar P}^{\vec J_0^{SS}\vec J_0^{ES}} & 0 & 0 & \bar{\bar P}^{\vec J_0^{SS}\vec M_0^{ES}}\\ \bar{\bar P}^{\vec J_0^{ES}\vec J_0^{SL}} & \bar{\bar P}^{\vec J_0^{ES}\vec J_0^{SS}} & \bar{\bar P}^{\vec J_0^{ES}\vec J_0^{ES}} & 0 & 0 & \bar{\bar P}^{\vec J_0^{ES}\vec M_0^{ES}}\\ 0 & 0 & 0 & 0 & 0 & 0\\ 0 & 0 & 0 & 0 & 0 & 0\\ \bar{\bar P}^{\vec M_0^{ES}\vec J_0^{SL}} & \bar{\bar P}^{\vec M_0^{ES}\vec J_0^{SS}} & \bar{\bar P}^{\vec M_0^{ES}\vec J_0^{ES}} & 0 & 0 & \bar{\bar P}^{\vec M_0^{ES}\vec M_0^{ES}} \end{bmatrix} \tag{65.1}$$

$$\bar a_{met-mat\ sys}^{\{\vec J_0^{SL},\vec J_0^{SS},\vec J_0^{ES},\vec J_\cap^{SL},\vec J_\cap^{SS},\vec M_0^{ES}\}} = \begin{bmatrix}\bar a^{\vec J_0^{SL}}\\ \bar a^{\vec J_0^{SS}}\\ \bar a^{\vec J_0^{ES}}\\ \bar a^{\vec J_\cap^{SL}}\\ \bar a^{\vec J_\cap^{SS}}\\ \bar a^{\vec M_0^{ES}}\end{bmatrix} \tag{65.2}$$

in which the elements of various submatrices are as follows:

$$p_{\xi\zeta}^{\vec J_0^Y\vec J_0^Z} = j\omega\mu_0\ (1/2)\left\langle \vec b_\xi^{\vec J_0^Y}\ ,\ \mathcal{L}_0\left(\vec b_\zeta^{\vec J_0^Z}\right)\right\rangle_{L_0^{met}\cup S_0^{met}\cup\partial V_0^{met}} \tag{66.1}$$

$$p_{\xi\zeta}^{\vec J_0^Y\vec J_0^{ES}} = -\ j\omega\mu_0\ (1/2)\left\langle \vec b_\xi^{\vec J_0^Y}\ ,\ \mathcal{L}_0\left(\vec b_\zeta^{\vec J_0^{ES}}\right)\right\rangle_{L_0^{met}\cup S_0^{met}\cup\partial V_0^{met}} \tag{66.2}$$

$$p_{\xi\zeta}^{\vec J_0^Y\vec M_0^{ES}} = -\ (1/2)\left\langle \vec b_\xi^{\vec J_0^Y}\ ,\ \mathcal{K}_0\left(\vec b_\zeta^{\vec M_0^{ES}}\right)\right\rangle_{L_0^{met}\cup S_0^{met}\cup\partial V_0^{met}} \tag{66.3}$$

$$p_{\xi\zeta}^{\vec J_0^{ES}\vec J_0^Z} = -\ j\omega\mu_0\ (1/2)\left\langle \vec b_\xi^{\vec J_0^{ES}}\ ,\ \mathcal{L}_0\left(\vec b_\zeta^{\vec J_0^Z}\right)\right\rangle_{\partial V_0^{mat}} \tag{66.4}$$

$$p_{\xi\zeta}^{\vec{J}_0^{ES}\vec{J}_0^{ES}} = j\omega\mu_0 \, (1/2)\left\langle \vec{b}_\xi^{\vec{J}_0^{ES}} , \mathcal{L}_0\left(\vec{b}_\zeta^{\vec{J}_0^{ES}}\right)\right\rangle_{\partial V_0^{mat}} \tag{66.5}$$

$$p_{\xi\zeta}^{\vec{J}_0^{ES}\vec{M}_0^{ES}} = (1/2)\left\langle \vec{b}_\xi^{\vec{J}_0^{ES}} , \mathcal{K}_0\left(\vec{b}_\zeta^{\vec{M}_0^{ES}}\right)\right\rangle_{\partial V_{0-}^{mat}} \tag{66.6}$$

$$p_{\xi\zeta}^{\vec{M}_0^{ES}\vec{M}_0^{ES}} = j\omega\varepsilon_0 \, (1/2)\left\langle \vec{b}_\xi^{\vec{M}_0^{ES}} , \mathcal{L}_0\left(\vec{b}_\zeta^{\vec{M}_0^{ES}}\right)\right\rangle_{\partial V_0^{mat}} \tag{66.7}$$

$$p_{\xi\zeta}^{\vec{M}_0^{ES}\vec{J}_0^{Z}} = (1/2)\left\langle \vec{b}_\xi^{\vec{M}_0^{ES}} , \mathcal{K}_0\left(\vec{b}_\zeta^{\vec{J}_0^{Z}}\right)\right\rangle_{\partial V_0^{mat}} \tag{66.8}$$

$$p_{\xi\zeta}^{\vec{M}_0^{ES}\vec{J}_0^{ES}} = - (1/2)\left\langle \vec{b}_\xi^{\vec{M}_0^{ES}} , \mathcal{K}_0\left(\vec{b}_\zeta^{\vec{J}_0^{ES}}\right)\right\rangle_{\partial V_{0-}^{mat}} \tag{66.9}$$

for any $Y,Z = SL,SS$, where the subscript " $-$ " used in integral domain $\partial V_{0-}^{mat}$ is to emphasize that the integral is done on the internal surface of boundary $\partial V_0^{mat}$.

**To relate equivalent electric and magnetic currents in expansion vector space**

It has been pointed out in [13] that: the equivalent electric and magnetic currents depend on each other, and it is an indispensable step for CM theory to relate equivalent electric and magnetic currents; if the equivalent electric and magnetic currents are not properly related to each other, some spurious modes will be generated. In the following parts of this subsection, the transformations between equivalent electric and magnetic currents are established by employing formulation (20) and artificial extinction theorem (45').

The equivalent electric current and equivalent magnetic current on material boundary satisfy the following relations:

$$\partial V_0^{mat} : \vec{J}_0^{ES}\times\hat{n}_{\rightarrow mat} = \left\{\left[\tilde{\bar{G}}_{sys}^{JH} * \vec{J}_\cap^{SL}\right]_{L_\cap^{met}}^{\tan} + \left[\tilde{\bar{G}}_{sys}^{JH} * \vec{J}_\cap^{SS}\right]_{S_\cap^{met}\cup\partial V_\cap^{met}}^{\tan} + \left[\tilde{\bar{G}}_{sys}^{JH} * \vec{J}_0^{ES} + \tilde{\bar{G}}_{sys}^{MH} * \vec{M}_0^{ES}\right]_{\partial V_0^{mat}}^{\tan}\right\} \tag{67.1}$$

$$\partial V_0^{mat} : \hat{n}_{\rightarrow mat}\times\vec{M}_0^{ES} = \left\{\left[\tilde{\bar{G}}_{sys}^{JE} * \vec{J}_\cap^{SL}\right]_{L_\cap^{met}}^{\tan} + \left[\tilde{\bar{G}}_{sys}^{JE} * \vec{J}_\cap^{SS}\right]_{S_\cap^{met}\cup\partial V_\cap^{met}}^{\tan} + \left[\tilde{\bar{G}}_{sys}^{JE} * \vec{J}_0^{ES} + \tilde{\bar{G}}_{sys}^{ME} * \vec{M}_0^{ES}\right]_{\partial V_0^{mat}}^{\tan}\right\} \tag{67.2}$$

as illustrated in (20) and (45'). If the (69.2) is tested by using basis functions set $\{\vec{b}_\xi^{\vec{J}_0^{ES}}\}$, then the expansion vector $\bar{a}^{\vec{J}_0^{ES}}$ can be expressed in terms of other expansion vectors as the following transformation:

$$\bar{a}^{\vec{J}_0^{ES}} = \bar{\bar{T}}_{met-mat\ sys}^{\{\vec{J}_\cap^{SL},\vec{J}_\cap^{SS},\vec{M}_0^{ES}\}\rightarrow\vec{J}_0^{ES}} \cdot \begin{bmatrix} \bar{a}^{\vec{J}_\cap^{SL}} \\ \bar{a}^{\vec{J}_\cap^{SS}} \\ \bar{a}^{\vec{M}_0^{ES}} \end{bmatrix} \tag{68}$$

where

$$\bar{\bar{T}}_{met-mat\ sys}^{\{\vec{J}_\cap^{SL},\vec{J}_\cap^{SS},\vec{M}_0^{ES}\}\rightarrow\vec{J}_0^{ES}} = \left(\bar{\bar{\Phi}}^{\vec{b}^{\vec{J}_0^{ES}}\vec{J}_0^{ES}}\right)^{-1}\cdot\left[\bar{\bar{\Phi}}^{\vec{b}^{\vec{J}_0^{ES}}\vec{J}_\cap^{SL}} \quad \bar{\bar{\Phi}}^{\vec{b}^{\vec{J}_0^{ES}}\vec{J}_\cap^{SS}} \quad \bar{\bar{\Phi}}^{\vec{b}^{\vec{J}_0^{ES}}\vec{M}_0^{ES}}\right] \tag{69}$$

in which the superscript " $-1$ " represents the inverse of matrix, and the elements of various submatrices are as follows:

$$\phi_{\xi\zeta}^{\vec{b}^{\vec{J}_0^{ES}}\vec{J}_0^{ES}} = \left\langle \vec{b}_\xi^{\vec{J}_0^{ES}} , \left[\tilde{\bar{G}}_{sys}^{JE} * \vec{b}_\zeta^{\vec{J}_0^{ES}}\right]_{\partial V_0^{mat}}\right\rangle_{\partial V_{0-}^{mat}} \tag{70.1}$$

$$\phi_{\xi\zeta}^{\vec{b}^{\vec{J}_0^{ES}}\vec{J}_\cap^{SL}} = -\left\langle \vec{b}_\xi^{\vec{J}_0^{ES}} , \left[\tilde{\bar{G}}_{sys}^{JE} * \vec{b}_\zeta^{\vec{J}_\cap^{SL}}\right]_{L_\cap^{met}}\right\rangle_{\partial V_{0-}^{mat}} \tag{70.2}$$

$$\phi_{\xi\zeta}^{\vec{b}^{\vec{J}_0^{ES}}\vec{J}_\cap^{SS}} = -\left\langle \vec{b}_\xi^{\vec{J}_0^{ES}} , \left[\tilde{\bar{G}}_{sys}^{JE} * \vec{b}_\zeta^{\vec{J}_\cap^{SS}}\right]_{S_\cap^{met}\cup\partial V_\cap^{met}}\right\rangle_{\partial V_{0-}^{mat}} \tag{70.3}$$

$$\phi_{\xi\zeta}^{\vec{b}^{\vec{J}_0^{ES}}\vec{M}_0^{ES}} = \left\langle \vec{b}_\xi^{\vec{J}_0^{ES}} , \hat{n}_{\rightarrow mat}\times\vec{b}_\zeta^{\vec{M}_0^{ES}} - \left[\tilde{\bar{G}}_{sys}^{ME} * \vec{b}_\zeta^{\vec{M}_0^{ES}}\right]_{\partial V_0^{mat}}\right\rangle_{\partial V_{0-}^{mat}} \tag{70.4}$$

where the subscript " $-$ " used in integral domain $\partial V_{0-}^{mat}$ is to emphasize that the integral is done on the internal surface of boundary $\partial V_0^{mat}$.

Inserting (68) into (64), (64) becomes the following form:

$$P_{met-mat\ sys}^{Harrington} = \left(\bar{a}_{met-mat\ sys}^{\{\vec{J}_0^{SL},\vec{J}_0^{SS},\vec{J}_\cap^{SL},\vec{J}_\cap^{SS},\vec{M}_0^{ES}\}}\right)^H \cdot \bar{\bar{P}}_{met-mat\ sys}^{\{\vec{J}_0^{SL},\vec{J}_0^{SS},\vec{J}_\cap^{SL},\vec{J}_\cap^{SS},\vec{M}_0^{ES}\}} \cdot \bar{a}_{met-mat\ sys}^{\{\vec{J}_0^{SL},\vec{J}_0^{SS},\vec{J}_\cap^{SL},\vec{J}_\cap^{SS},\vec{M}_0^{ES}\}} \tag{71}$$

where

$$\bar{\bar{P}}_{met-mat\ sys}^{\{\vec{J}_0^{SL},\vec{J}_0^{SS},\vec{J}_\cap^{SL},\vec{J}_\cap^{SS},\vec{M}_0^{ES}\}} = \begin{bmatrix} \bar{\bar{I}} & 0 & 0 \\ 0 & \bar{\bar{I}} & 0 \\ 0 & 0 & \bar{\bar{T}}_{met-mat\ sys}^{\{\vec{J}_\cap^{SL},\vec{J}_\cap^{SS},\vec{M}_0^{ES}\}\rightarrow\vec{J}_0^{ES}} \\ 0 & 0 & \bar{\bar{I}} \end{bmatrix}^H \cdot \bar{\bar{P}}_{met-mat\ sys}^{\{\vec{J}_0^{SL},\vec{J}_0^{SS},\vec{J}_\cap^{SL},\vec{J}_\cap^{SS},\vec{M}_0^{ES}\}} \cdot \begin{bmatrix} \bar{\bar{I}} & 0 & 0 \\ 0 & \bar{\bar{I}} & 0 \\ 0 & 0 & \bar{\bar{T}}_{met-mat\ sys}^{\{\vec{J}_\cap^{SL},\vec{J}_\cap^{SS},\vec{M}_0^{ES}\}\rightarrow\vec{J}_0^{ES}} \\ 0 & 0 & \bar{\bar{I}} \end{bmatrix} \tag{72.1}$$

$$\bar{a}_{met-mat\ sys}^{\{\vec{J}_0^{SL},\vec{J}_0^{SS},\vec{J}_\cap^{SL},\vec{J}_\cap^{SS},\vec{M}_0^{ES}\}} = \begin{bmatrix} \bar{a}^{\vec{J}_0^{SL}} \\ \bar{a}^{\vec{J}_0^{SS}} \\ \bar{a}^{\{\vec{J}_\cap^{SL},\vec{J}_\cap^{SS},\vec{M}_0^{ES}\}} \end{bmatrix}. \tag{72.2}$$

In (72.2),

$$\bar{a}^{\{\vec{J}_\cap^{SL},\vec{J}_\cap^{SS},\vec{M}_0^{ES}\}} = \begin{bmatrix} \bar{a}^{\vec{J}_\cap^{SL}} \\ \bar{a}^{\vec{J}_\cap^{SS}} \\ \bar{a}^{\vec{M}_0^{ES}} \end{bmatrix}. \tag{72.3}$$

Similarly to establishing (68) by testing (67.2) with $\{\vec{b}_\xi^{\vec{J}_0^{ES}}\}$, the following transformation can be easily established:

$$\bar{a}^{\vec{M}_0^{ES}} = \bar{\bar{T}}_{met-mat\ sys}^{\{\vec{J}_0^{ES},\vec{J}_\cap^{SL},\vec{J}_\cap^{SS}\}\rightarrow\vec{M}_0^{ES}} \cdot \begin{bmatrix} \bar{a}^{\vec{J}_0^{ES}} \\ \bar{a}^{\vec{J}_\cap^{SL}} \\ \bar{a}^{\vec{J}_\cap^{SS}} \end{bmatrix} \tag{73}$$

by testing (67.1) with basis functions set $\{\vec{b}_\xi^{\vec{M}_0^{ES}}\}$. Inserting (73) into (64), (64) becomes the following form:

$$P_{met-mat\ sys}^{Harrington} = \left(\bar{a}_{met-mat\ sys}^{\{\vec{J}_0^{SL},\vec{J}_0^{SS},\vec{J}_0^{ES},\vec{J}_\cap^{SL},\vec{J}_\cap^{SS}\}}\right)^H \cdot \bar{\bar{P}}_{met-mat\ sys}^{\{\vec{J}_0^{SL},\vec{J}_0^{SS},\vec{J}_0^{ES},\vec{J}_\cap^{SL},\vec{J}_\cap^{SS}\}} \cdot \bar{a}_{met-mat\ sys}^{\{\vec{J}_0^{SL},\vec{J}_0^{SS},\vec{J}_0^{ES},\vec{J}_\cap^{SL},\vec{J}_\cap^{SS}\}} \tag{74}$$

where

$$\overline{\overline{P}}_{met-mat\ sys}^{\left\{\vec{J}_{0}^{SL},\vec{J}_{0}^{SS},\vec{J}_{0}^{ES},\vec{J}_{\cap}^{SL},\vec{J}_{\cap}^{SS}\right\}}=\begin{bmatrix}\overline{\overline{I}} & 0 & 0\\ 0 & \overline{\overline{I}} & 0\\ 0 & 0 & \overline{\overline{I}}\\ 0 & 0 & \overline{\overline{T}}_{met-mat\ sys}^{\left\{\vec{J}_{0}^{ES},\vec{J}_{\cap}^{SL},\vec{J}_{\cap}^{SS}\right\}\rightarrow\vec{M}_{0}^{ES}}\end{bmatrix}^{H}\cdot\overline{\overline{P}}_{met-mat\ sys}^{\left\{\vec{J}_{0}^{SL},\vec{J}_{0}^{SS},\vec{J}_{0}^{ES},\vec{J}_{\cap}^{SL},\vec{J}_{\cap}^{SS},\vec{M}_{0}^{ES}\right\}}\cdot\begin{bmatrix}\overline{\overline{I}} & 0 & 0\\ 0 & \overline{\overline{I}} & 0\\ 0 & 0 & \overline{\overline{I}}\\ 0 & 0 & \overline{\overline{T}}_{met-mat\ sys}^{\left\{\vec{J}_{0}^{ES},\vec{J}_{\cap}^{SL},\vec{J}_{\cap}^{SS}\right\}\rightarrow\vec{M}_{0}^{ES}}\end{bmatrix}\tag{75.1}$$

$$\overline{a}_{met-mat\ sys}^{\left\{\vec{J}_{0}^{SL},\vec{J}_{0}^{SS},\vec{J}_{0}^{ES},\vec{J}_{\cap}^{SL},\vec{J}_{\cap}^{SS}\right\}}=\begin{bmatrix}\overline{a}^{\vec{J}_{0}^{SL}}\\ \overline{a}^{\vec{J}_{0}^{SS}}\\ \overline{a}^{\left\{\vec{J}_{0}^{ES},\vec{J}_{\cap}^{SL},\vec{J}_{\cap}^{SS}\right\}}\end{bmatrix}.\tag{75.2}$$

In (75.2),

$$\overline{a}^{\left\{\vec{J}_{0}^{ES},\vec{J}_{\cap}^{SL},\vec{J}_{\cap}^{SS}\right\}}=\begin{bmatrix}\overline{a}^{\vec{J}_{0}^{ES}}\\ \overline{a}^{\vec{J}_{\cap}^{SL}}\\ \overline{a}^{\vec{J}_{\cap}^{SS}}\end{bmatrix}.\tag{75.3}$$

### *D. Harrington's CM orthogonalizing operator (53)*

Taking matrix form (71) as an example, the Harrington's CM of metal-material combined system is constructed as below.

The power matrix $\overline{\overline{P}}_{met-mat\ sys}^{\left\{\vec{J}_{0}^{SL},\vec{J}_{0}^{SS},\vec{J}_{\cap}^{SL},\vec{J}_{\cap}^{SS},\vec{M}_{0}^{ES}\right\}}$ can be decomposed as

$$\overline{\overline{P}}_{met-mat\ sys}^{\left\{\vec{J}_{0}^{SL},\vec{J}_{0}^{SS},\vec{J}_{\cap}^{SL},\vec{J}_{\cap}^{SS},\vec{M}_{0}^{ES}\right\}}=\overline{\overline{P}}_{met-mat\ sys;+}^{\left\{\vec{J}_{0}^{SL},\vec{J}_{0}^{SS},\vec{J}_{\cap}^{SL},\vec{J}_{\cap}^{SS},\vec{M}_{0}^{ES}\right\}}+j\,\overline{\overline{P}}_{met-mat\ sys;-}^{\left\{\vec{J}_{0}^{SL},\vec{J}_{0}^{SS},\vec{J}_{\cap}^{SL},\vec{J}_{\cap}^{SS},\vec{M}_{0}^{ES}\right\}}\tag{76}$$

where [13]

$$\overline{\overline{P}}_{met-mat\ sys;+}^{\left\{\vec{J}_{0}^{SL},\vec{J}_{0}^{SS},\vec{J}_{\cap}^{SL},\vec{J}_{\cap}^{SS},\vec{M}_{0}^{ES}\right\}}=\frac{1}{2}\left[\overline{\overline{P}}_{met-mat\ sys}^{\left\{\vec{J}_{0}^{SL},\vec{J}_{0}^{SS},\vec{J}_{\cap}^{SL},\vec{J}_{\cap}^{SS},\vec{M}_{0}^{ES}\right\}}+\left(\overline{\overline{P}}_{met-mat\ sys}^{\left\{\vec{J}_{0}^{SL},\vec{J}_{0}^{SS},\vec{J}_{\cap}^{SL},\vec{J}_{\cap}^{SS},\vec{M}_{0}^{ES}\right\}}\right)^{H}\right]\tag{77.1}$$

$$\overline{\overline{P}}_{met-mat\ sys;-}^{\left\{\vec{J}_{0}^{SL},\vec{J}_{0}^{SS},\vec{J}_{\cap}^{SL},\vec{J}_{\cap}^{SS},\vec{M}_{0}^{ES}\right\}}=\frac{1}{2j}\left[\overline{\overline{P}}_{met-mat\ sys}^{\left\{\vec{J}_{0}^{SL},\vec{J}_{0}^{SS},\vec{J}_{\cap}^{SL},\vec{J}_{\cap}^{SS},\vec{M}_{0}^{ES}\right\}}-\left(\overline{\overline{P}}_{met-mat\ sys}^{\left\{\vec{J}_{0}^{SL},\vec{J}_{0}^{SS},\vec{J}_{\cap}^{SL},\vec{J}_{\cap}^{SS},\vec{M}_{0}^{ES}\right\}}\right)^{H}\right].\tag{77.2}$$

Based on Harrington's classical method [5], [11], [12], the CM can be obtained by solving characteristic equation

$$\begin{aligned}&\overline{\overline{P}}_{met-mat\ sys;-}^{\left\{\vec{J}_{0}^{SL},\vec{J}_{0}^{SS},\vec{J}_{\cap}^{SL},\vec{J}_{\cap}^{SS},\vec{M}_{0}^{ES}\right\}}\cdot\overline{a}_{met-mat\ sys;\xi}^{\left\{\vec{J}_{0}^{SL},\vec{J}_{0}^{SS},\vec{J}_{\cap}^{SL},\vec{J}_{\cap}^{SS},\vec{M}_{0}^{ES}\right\}}\\&=\lambda_{met-mat\ sys;\xi}\,\overline{\overline{P}}_{met-mat\ sys;+}^{\left\{\vec{J}_{0}^{SL},\vec{J}_{0}^{SS},\vec{J}_{\cap}^{SL},\vec{J}_{\cap}^{SS},\vec{M}_{0}^{ES}\right\}}\cdot\overline{a}_{met-mat\ sys;\xi}^{\left\{\vec{J}_{0}^{SL},\vec{J}_{0}^{SS},\vec{J}_{\cap}^{SL},\vec{J}_{\cap}^{SS},\vec{M}_{0}^{ES}\right\}}.\end{aligned}\tag{78}$$

## VII. APPLICATION OF GFHF: TO CONSTRUCT THE ELECTROMAGNETIC-POWER-BASED CM OF METAL-MATERIAL COMBINED SYSTEM

In papers [15]-[17], the electromagnetic-power-based (EMP-based) CM of metal-material combined system was constructed, and the material sub-system was restricted to being homogeneous and isotropic. In this section, some results obtained in [15]-[17] are generalized to the metal-material combined system whose material sub-system is inhomogeneous, anisotropic, LOSSLESS, and NON-MAGNETIC. At the same time, a new EMP-based CM set, optimally radiative intrinsically resonant CM (OptRadIntResCM) set, is proposed here.

### *A. Various powers*

Based on the conclusions given in [13] and [18] and the above Sec. VII, when the permeability of a EM system is $\mu_0$, the input power $P_{met-mat\ sys}^{inp}$ (i.e. the power done by incident fields on scattering currents) equals to Harrington's power $P_{met-mat\ sys}^{Harrington}$.

Then, taking the matrix form (71) as an example, the matrix form of $P_{met-mat\ sys}^{inp}$ is as follows:

$$P_{met-mat\ sys}^{inp}=\left(\overline{a}_{met-mat\ sys}^{\left\{\vec{J}_{0}^{SL},\vec{J}_{0}^{SS},\vec{J}_{\cap}^{SL},\vec{J}_{\cap}^{SS},\vec{M}_{0}^{ES}\right\}}\right)^{H}\cdot\overline{\overline{P}}_{met-mat\ sys}^{\left\{\vec{J}_{0}^{SL},\vec{J}_{0}^{SS},\vec{J}_{\cap}^{SL},\vec{J}_{\cap}^{SS},\vec{M}_{0}^{ES}\right\}}\cdot\overline{a}_{met-mat\ sys}^{\left\{\vec{J}_{0}^{SL},\vec{J}_{0}^{SS},\vec{J}_{\cap}^{SL},\vec{J}_{\cap}^{SS},\vec{M}_{0}^{ES}\right\}}\tag{79}$$

and the radiated and reactive powers are as follows:

$$\begin{aligned}P_{met-mat\ sys}^{rad}&=\mathrm{Re}\left\{P_{met-mat\ sys}^{inp}\right\}\\&=\left(\overline{a}_{met-mat\ sys}^{\left\{\vec{J}_{0}^{SL},\vec{J}_{0}^{SS},\vec{J}_{\cap}^{SL},\vec{J}_{\cap}^{SS},\vec{M}_{0}^{ES}\right\}}\right)^{H}\cdot\overline{\overline{P}}_{met-mat\ sys;+}^{\left\{\vec{J}_{0}^{SL},\vec{J}_{0}^{SS},\vec{J}_{\cap}^{SL},\vec{J}_{\cap}^{SS},\vec{M}_{0}^{ES}\right\}}\cdot\overline{a}_{met-mat\ sys}^{\left\{\vec{J}_{0}^{SL},\vec{J}_{0}^{SS},\vec{J}_{\cap}^{SL},\vec{J}_{\cap}^{SS},\vec{M}_{0}^{ES}\right\}}\end{aligned}\tag{80.1}$$

$$\begin{aligned}P_{met-mat\ sys}^{react}&=\mathrm{Im}\left\{P_{met-mat\ sys}^{inp}\right\}\\&=\left(\overline{a}_{met-mat\ sys}^{\left\{\vec{J}_{0}^{SL},\vec{J}_{0}^{SS},\vec{J}_{\cap}^{SL},\vec{J}_{\cap}^{SS},\vec{M}_{0}^{ES}\right\}}\right)^{H}\cdot\overline{\overline{P}}_{met-mat\ sys;-}^{\left\{\vec{J}_{0}^{SL},\vec{J}_{0}^{SS},\vec{J}_{\cap}^{SL},\vec{J}_{\cap}^{SS},\vec{M}_{0}^{ES}\right\}}\cdot\overline{a}_{met-mat\ sys}^{\left\{\vec{J}_{0}^{SL},\vec{J}_{0}^{SS},\vec{J}_{\cap}^{SL},\vec{J}_{\cap}^{SS},\vec{M}_{0}^{ES}\right\}}\end{aligned}\tag{80.2}$$

based on the results given in above Sec. VII.

### *B. Optimally radiative intrinsically resonant CM*

In papers [18] and [19], the radiated power CM (RadCM) set was introduced, and it has ability to optimize radiated power. However, the RadCM set cannot guarantee the orthogonality of modal reactive powers. In paper [17], the intrinsically resonant CM (IntResCM) set was introduced, and it constitute the basis of whole intrinsic resonance space as explained in [20]. However, the IntResCM set cannot guarantee that the IntResCMs can efficiently radiate EM energies as explained in paper [20].

Following the ideas of papers [17]-[20], a new EMP-based CM set, optimally radiative intrinsically resonant CM (OptRadIntResCM) set, is introduced in this section. The OptRadIntResCMs constitute a basis of whole intrinsic resonance space [20], and at the same time [the most efficiently radiative] intrinsically resonant modes can be found in the OptRadIntResCM set.

**Intrinsically resonant CMs**

Based on paper [17], the IntResCMs can be obtained by solving the following equation:

$$\overline{\overline{P}}_{met-mat\ sys;-}^{\left\{\vec{J}_{0}^{SL},\vec{J}_{0}^{SS},\vec{J}_{\cap}^{SL},\vec{J}_{\cap}^{SS},\vec{M}_{0}^{ES}\right\}}\cdot\overline{a}_{met-mat\ sys;int\ res;\xi}^{\left\{\vec{J}_{0}^{SL},\vec{J}_{0}^{SS},\vec{J}_{\cap}^{SL},\vec{J}_{\cap}^{SS},\vec{M}_{0}^{ES}\right\}}=0.\tag{81}$$

**From whole modal space to intrinsic resonance space**

If we obtain $\Xi_{met-mat\ sys;int\ res}^{\left\{\vec{J}_{0}^{SL},\vec{J}_{0}^{SS},\vec{J}_{\cap}^{SL},\vec{J}_{\cap}^{SS},\vec{M}_{0}^{ES}\right\}}$ IntResCMs, then any intrinsically resonant mode $\overline{a}_{met-mat\ sys;int\ res}^{\left\{\vec{J}_{0}^{SL},\vec{J}_{0}^{SS},\vec{J}_{\cap}^{SL},\vec{J}_{\cap}^{SS},\vec{M}_{0}^{ES}\right\}}$ can be expanded in terms of IntResCM set as follows: [20]

$$\begin{aligned}\overline{a}_{met-mat\ sys\,;\,int\ res}^{\left\{\vec{J}_{0}^{SL},\vec{J}_{0}^{SS},\vec{J}_{\cap}^{SL},\vec{J}_{\cap}^{SS},\vec{M}_{0}^{ES}\right\}}&=\sum_{\xi=1}^{\Xi_{met-mat\ sys;int\ res}^{\left\{\vec{J}_{0}^{SL},\vec{J}_{0}^{SS},\vec{J}_{\cap}^{SL},\vec{J}_{\cap}^{SS},\vec{M}_{0}^{ES}\right\}}}\alpha_{met-mat\ sys\,;\,int\ res\,;\,\xi}^{\left\{\vec{J}_{0}^{SL},\vec{J}_{0}^{SS},\vec{J}_{\cap}^{SL},\vec{J}_{\cap}^{SS},\vec{M}_{0}^{ES}\right\}}\overline{a}_{met-mat\ sys\,;\,int\ res\,;\,\xi}^{\left\{\vec{J}_{0}^{SL},\vec{J}_{0}^{SS},\vec{J}_{\cap}^{SL},\vec{J}_{\cap}^{SS},\vec{M}_{0}^{ES}\right\}}\\&=\overline{\overline{A}}_{met-mat\ sys\,;\,int\ res}^{\left\{\vec{J}_{0}^{SL},\vec{J}_{0}^{SS},\vec{J}_{\cap}^{SL},\vec{J}_{\cap}^{SS},\vec{M}_{0}^{ES}\right\}}\cdot\overline{\alpha}_{met-mat\ sys\,;\,int\ res}^{\left\{\vec{J}_{0}^{SL},\vec{J}_{0}^{SS},\vec{J}_{\cap}^{SL},\vec{J}_{\cap}^{SS},\vec{M}_{0}^{ES}\right\}}\end{aligned}\tag{82}$$

where

$$\begin{aligned}&\overline{\overline{A}}{}^{\{\vec{J}_0^{SL},\vec{J}_0^{SS},\vec{J}_\cap^{SL},\vec{J}_\cap^{SS},\vec{M}_0^{ES}\}}_{\text{met-mat sys ; int res}}\\ =&\begin{bmatrix}\bar{a}^{\{\vec{J}_0^{SL},\vec{J}_0^{SS},\vec{J}_\cap^{SL},\vec{J}_\cap^{SS},\vec{M}_0^{ES}\}}_{\text{met-mat sys ; int res ; 1}} & \cdots & \bar{a}^{\{\vec{J}_0^{SL},\vec{J}_0^{SS},\vec{J}_\cap^{SL},\vec{J}_\cap^{SS},\vec{M}_0^{ES}\}}_{\text{met-mat sys ; int res ; }\Xi^{\{\vec{J}_0^{SL},\vec{J}_0^{SS},\vec{J}_\cap^{SL},\vec{J}_\cap^{SS},\vec{M}_0^{ES}\}}_{\text{met-mat sys ; int res}}}\end{bmatrix}\end{aligned} \tag{83.1}$$

$$\begin{aligned}&\bar{\alpha}^{\{\vec{J}_0^{SL},\vec{J}_0^{SS},\vec{J}_\cap^{SL},\vec{J}_\cap^{SS},\vec{M}_0^{ES}\}}_{\text{met-mat sys ; int res}}\\ =&\begin{bmatrix}\alpha^{\{\vec{J}_0^{SL},\vec{J}_0^{SS},\vec{J}_\cap^{SL},\vec{J}_\cap^{SS},\vec{M}_0^{ES}\}}_{\text{met-mat sys ; int res ; 1}} & \cdots & \alpha^{\{\vec{J}_0^{SL},\vec{J}_0^{SS},\vec{J}_\cap^{SL},\vec{J}_\cap^{SS},\vec{M}_0^{ES}\}}_{\text{met-mat sys ; int res ; }\Xi^{\{\vec{J}_0^{SL},\vec{J}_0^{SS},\vec{J}_\cap^{SL},\vec{J}_\cap^{SS},\vec{M}_0^{ES}\}}_{\text{met-mat sys ; int res}}}\end{bmatrix}^T .\end{aligned} \tag{83.2}$$

Inserting the above (82) into (80.1), we have that

$$P^{rad}_{int\ res} = \left(\bar{\alpha}^{\{\vec{J}_0^{SL},\vec{J}_0^{SS},\vec{J}_\cap^{SL},\vec{J}_\cap^{SS},\vec{M}_0^{ES}\}}_{\text{met-mat sys ; int res}}\right)^H \cdot \overline{\overline{P}}{}^{\{\vec{J}_0^{SL},\vec{J}_0^{SS},\vec{J}_\cap^{SL},\vec{J}_\cap^{SS},\vec{M}_0^{ES}\}}_{\text{met-mat sys ; + ; int res}} \cdot \bar{\alpha}^{\{\vec{J}_0^{SL},\vec{J}_0^{SS},\vec{J}_\cap^{SL},\vec{J}_\cap^{SS},\vec{M}_0^{ES}\}}_{\text{met-mat sys ; int res}} \tag{84}$$

where

$$\begin{aligned}&\overline{\overline{P}}{}^{\{\vec{J}_0^{SL},\vec{J}_0^{SS},\vec{J}_\cap^{SL},\vec{J}_\cap^{SS},\vec{M}_0^{ES}\}}_{\text{met-mat sys ; + ; int res}}\\ =&\left(\overline{\overline{A}}{}^{\{\vec{J}_0^{SL},\vec{J}_0^{SS},\vec{J}_\cap^{SL},\vec{J}_\cap^{SS},\vec{M}_0^{ES}\}}_{\text{met-mat sys ; int res}}\right)^H \cdot \overline{\overline{P}}{}^{\{\vec{J}_0^{SL},\vec{J}_0^{SS},\vec{J}_\cap^{SL},\vec{J}_\cap^{SS},\vec{M}_0^{ES}\}}_{\text{met-mat sys ; +}} \cdot \overline{\overline{A}}{}^{\{\vec{J}_0^{SL},\vec{J}_0^{SS},\vec{J}_\cap^{SL},\vec{J}_\cap^{SS},\vec{M}_0^{ES}\}}_{\text{met-mat sys ; int res}} .\end{aligned} \tag{85}$$

**Normalized radiated powers of intrinsically resonant modes**

Following the normalization way proposed in papers [18] and [19], the normalized $P^{rad}_{int\ res}$ is as follows:

$$\begin{aligned}\tilde{P}^{rad}_{int\ res} &= \frac{P^{rad}_{int\ res}}{\left(\bar{\alpha}^{\{\vec{J}_0^{SL},\vec{J}_0^{SS},\vec{J}_\cap^{SL},\vec{J}_\cap^{SS},\vec{M}_0^{ES}\}}_{\text{met-mat sys ; int res}}\right)^H \cdot \overline{\overline{C}}{}^{\{\vec{J}_0^{SL},\vec{J}_0^{SS},\vec{J}_\cap^{SL},\vec{J}_\cap^{SS},\vec{M}_0^{ES}\}}_{\text{met-mat sys ; int res}} \cdot \bar{\alpha}^{\{\vec{J}_0^{SL},\vec{J}_0^{SS},\vec{J}_\cap^{SL},\vec{J}_\cap^{SS},\vec{M}_0^{ES}\}}_{\text{met-mat sys ; int res}}}\\ &= \frac{\left(\bar{\alpha}^{\{\vec{J}_0^{SL},\vec{J}_0^{SS},\vec{J}_\cap^{SL},\vec{J}_\cap^{SS},\vec{M}_0^{ES}\}}_{\text{met-mat sys ; int res}}\right)^H \cdot \overline{\overline{P}}{}^{\{\vec{J}_0^{SL},\vec{J}_0^{SS},\vec{J}_\cap^{SL},\vec{J}_\cap^{SS},\vec{M}_0^{ES}\}}_{\text{met-mat sys ; + ; int res}} \cdot \bar{\alpha}^{\{\vec{J}_0^{SL},\vec{J}_0^{SS},\vec{J}_\cap^{SL},\vec{J}_\cap^{SS},\vec{M}_0^{ES}\}}_{\text{met-mat sys ; int res}}}{\left(\bar{\alpha}^{\{\vec{J}_0^{SL},\vec{J}_0^{SS},\vec{J}_\cap^{SL},\vec{J}_\cap^{SS},\vec{M}_0^{ES}\}}_{\text{met-mat sys ; int res}}\right)^H \cdot \overline{\overline{C}}{}^{\{\vec{J}_0^{SL},\vec{J}_0^{SS},\vec{J}_\cap^{SL},\vec{J}_\cap^{SS},\vec{M}_0^{ES}\}}_{\text{met-mat sys ; int res}} \cdot \bar{\alpha}^{\{\vec{J}_0^{SL},\vec{J}_0^{SS},\vec{J}_\cap^{SL},\vec{J}_\cap^{SS},\vec{M}_0^{ES}\}}_{\text{met-mat sys ; int res}}}\end{aligned} \tag{86}$$

where

$$\begin{aligned}&\overline{\overline{C}}{}^{\{\vec{J}_0^{SL},\vec{J}_0^{SS},\vec{J}_\cap^{SL},\vec{J}_\cap^{SS},\vec{M}_0^{ES}\}}_{\text{met-mat sys ; int res}}\\ =&\left(\overline{\overline{A}}{}^{\{\vec{J}_0^{SL},\vec{J}_0^{SS},\vec{J}_\cap^{SL},\vec{J}_\cap^{SS},\vec{M}_0^{ES}\}}_{\text{met-mat sys ; int res}}\right)^H \cdot \overline{\overline{C}}{}^{\{\vec{J}_0^{SL},\vec{J}_0^{SS},\vec{J}_\cap^{SL},\vec{J}_\cap^{SS},\vec{M}_0^{ES}\}}_{\text{met-mat sys}} \cdot \overline{\overline{A}}{}^{\{\vec{J}_0^{SL},\vec{J}_0^{SS},\vec{J}_\cap^{SL},\vec{J}_\cap^{SS},\vec{M}_0^{ES}\}}_{\text{met-mat sys ; int res}}\end{aligned} \tag{87}$$

and

$$\overline{\overline{C}}{}^{\{\vec{J}_0^{SL},\vec{J}_0^{SS},\vec{J}_\cap^{SL},\vec{J}_\cap^{SS},\vec{M}_0^{ES}\}}_{\text{met-mat sys}} = \begin{bmatrix}\overline{\overline{J}}{}_0^{SL} & 0 & 0 & 0 & 0\\ 0 & \overline{\overline{J}}{}_0^{SS} & 0 & 0 & 0\\ 0 & 0 & \overline{\overline{J}}{}_\cap^{SL} & 0 & 0\\ 0 & 0 & 0 & \overline{\overline{J}}{}_\cap^{SS} & 0\\ 0 & 0 & 0 & 0 & \overline{\overline{M}}{}_0^{ES}\end{bmatrix} . \tag{88}$$

In (88), the elements of various sub-matrices are as follows:

$$j^{SL}_{0/\cap;\xi\zeta} = \frac{1}{2\left\|L^{met}_{0/\cap}\right\|}\left\langle \vec{b}_\xi^{\vec{J}^{SL}_{0/\cap}} , \vec{b}_\zeta^{\vec{J}^{SL}_{0/\cap}}\right\rangle_{L^{met}_{0/\cap}} \tag{89.1}$$

$$j^{SS}_{0/\cap;\xi\zeta} = \frac{1}{2}\left\langle \vec{b}_\xi^{\vec{J}^{SS}_{0/\cap}} , \vec{b}_\zeta^{\vec{J}^{SS}_{0/\cap}}\right\rangle_{S^{met}_{0/\cap}\cup\partial V^{met}_{0/\cap}} \tag{89.2}$$

$$m^{ES}_{0;\xi\zeta} = \frac{1}{2\eta_0^2}\left\langle \vec{b}_\xi^{\vec{M}^{ES}_0} , \vec{b}_\zeta^{\vec{M}^{ES}_0}\right\rangle_{\partial V^{mat}_0} \tag{89.3}$$

where the $\left\|L^{met}_{0/\cap}\right\|$ is the length of $L^{met}_{0/\cap}$, and the $\eta_0$ is wave impedance in vacuum.

**Optimally radiative intrinsically resonant CM**

Following the ideas of papers [18] and [19], the OptRadIntResCMs can be derived from solving the following generalized characteristic equation:

$$\begin{aligned}&\overline{\overline{P}}{}^{\{\vec{J}_0^{SL},\vec{J}_0^{SS},\vec{J}_\cap^{SL},\vec{J}_\cap^{SS},\vec{M}_0^{ES}\}}_{\text{met-mat sys ; + ; int res}} \cdot \bar{\alpha}^{\{\vec{J}_0^{SL},\vec{J}_0^{SS},\vec{J}_\cap^{SL},\vec{J}_\cap^{SS},\vec{M}_0^{ES}\}}_{\text{met-mat sys ; int res ; }\xi}\\ =& R^{rad}_{\text{met-mat sys ; int res ; }\xi}\ \overline{\overline{C}}{}^{\{\vec{J}_0^{SL},\vec{J}_0^{SS},\vec{J}_\cap^{SL},\vec{J}_\cap^{SS},\vec{M}_0^{ES}\}}_{\text{met-mat sys ; int res}} \cdot \bar{\alpha}^{\{\vec{J}_0^{SL},\vec{J}_0^{SS},\vec{J}_\cap^{SL},\vec{J}_\cap^{SS},\vec{M}_0^{ES}\}}_{\text{met-mat sys ; int res ; }\xi} .\end{aligned} \tag{90}$$

## VIII. Conclusions

In this paper, the formulations and conclusions given in our previous works, which focus on the EM system constructed by inhomogeneous anisotropic lossy material bodies, are generalized to metal-material combined EM system. The formulations appeared in both this paper and our previous works are formally unified, and the conclusions in both this paper and our previous works are consistent.

In previous our works, it is pointed out that the GFHF of material system is the mathematical expression of SEP rather than the mathematical expression of GHP. In this paper, it is found out that the GFHF of metal-material combined system is the mathematical expression of line-surface equivalence principle.

The values of GFHF are mainly manifested in that various fields are uniformly expressed in terms of an identical set of currents, and this feature is very valuable for many engineering applications, such as solving EM scattering problem and constructing CM set, and some typical applications are exhibited in this paper.

## Acknowledgement

This work is dedicated to my mother for her constant understanding, support, and encouragement.